\documentclass[%
 reprint,
 amsmath,amssymb,
 aps,
]{revtex4-2}

\usepackage{graphicx,xcolor}
\usepackage{dcolumn}
\usepackage{bm}
\usepackage{float}

\begin{document}

\preprint{APS/123-QED}

\title{Band Structure and Effective Properties of\\ One-Dimensional Thermoacoustic Bloch Waves}

\author{Haitian Hao}
\email[]{haoh@purdue.edu}
\affiliation{School of Mechanical Engineering, Purdue University, West Lafayette, Indiana 47907, USA}

\author{Carlo Scalo}
\email[]{scalo@purdue.edu}
\affiliation{School of Mechanical Engineering, Purdue University, West Lafayette, Indiana 47907, USA}

\author{Fabio Semperlotti}
\email[]{Corresponding Author: fsemperl@purdue.edu}
\affiliation{School of Mechanical Engineering, Purdue University, West Lafayette, Indiana 47907, USA}


\begin{abstract}

We investigate the dispersion characteristics and the effective properties of acoustic waves propagating in a one-dimensional duct equipped with periodic thermoacoustic coupling elements. Each coupling element consists in a classical thermoacoustic regenerator subject to a spatial temperature gradient. When acoustic waves pass through the regenerator, thermal-to-acoustic energy conversion takes place and can either amplify or attenuate the wave, depending on the direction of propagation of the wave. The presence of the spatial gradient naturally induces a loss of reciprocity. This study provides a comprehensive theoretical model as well as an in-depth numerical analysis of the band structure and of the propagation properties of this thermoacoustically-coupled, tunable, one-dimensional metamaterial. Among the most significant findings, it is shown that the acoustic metamaterial is capable of supporting non-reciprocal thermoacoustic Bloch waves that are associated with a particular form of unidirectional energy transport. Remarkably, the thermoacoustic coupling also allows achieving effective zero compressibility and zero refractive index that ultimately lead to the phase invariance of the propagating sound waves. This single zero effective property is also shown to have very interesting implications in the attainment of acoustic cloaking. 
\end{abstract}

\maketitle

\section{Introduction}
In recent years, the study of acoustic metamaterials has focused on the possibility to break reciprocity and on the resulting effects on the dispersion and propagation of sound \cite{Fleury14, Fleury15,Swinteck,Boechler,Liang09,Liang10}. In conventional acoustic waveguides (e.g. a hollow duct), sound waves are reciprocally transmitted between two points of the domain. Exciting the domain at a source $A$ and measuring its response at a point $B$ would yield the exact same response if the source and the observation point were inverted. However, this reciprocal wave transmission mechanism might not always be a desired feature. There are certain applications such as, medical imaging \cite{Zhu} or telecommunications devices \cite{Barzanjeh} whose performance can be significantly improved in presence of unidirectional sound transmission. For completeness, it is worth mentioning that unidirectional propagation has been observed and studied either in non-reciprocal or topological systems. While, under certain conditions, the effect of the two systems on the wave propagation characteristics might be conflated, the two systems lead to unidirectional propagation by means of very different mechanisms. Indeed, many topologically non-trivial systems are still reciprocal in nature \cite{Yang,Ni,TWLiu}.
Focusing on non-reciprocity, non-reciprocal waves have been achieved in a variety of systems leveraging, as an example, rotating fluids \cite{Fleury14}, active materials with spatiotemporal modulation \cite{Fleury15,Swinteck}, near zero refractive index materials \cite{LiY} and materials with strong nonlinearities \cite{Boechler,Liang09,Liang10}. Non-reciprocal propagation was also observed in thermoacoustically coupled systems consisting in torus-shaped thermoacoustic (TA) engines \cite{Yazaki,Swift88,Gupta,HaoJSV19}. It was this class of systems that later inspired the design of TA diodes \cite{Biwa16} and TA amplifiers \cite{Senga}. Despite the long and distinguished history of thermoacoustics and the more recent analysis of diodes and amplifiers, the systematic analysis and in-depth understanding of the dispersion and propagation properties in periodic TA waveguides have never been undertaken. 

The current study specifically addresses this latter point by presenting a comprehensive theoretical and numerical analysis of the dispersion and propagation properties of thermoacoustically coupled waves in one-dimensional ducts embedded with a periodic distribution of regenerators. In the following, we will refer to the acoustic waves supported by this type of waveguides as \textit{thermoacoustic Bloch waves}. This specific type of 1D waveguides can be seen as a form of semi-active acoustic metamaterial where thermo-acoustic energy conversion occurs periodically when the fluid passes through the evenly-spaced regenerators. The semi-active nature of the system is due to the fact that energy is either provided or extracted from the acoustic wave as a consequence of the imposed static thermal gradient on the regenerators.
The regenerator (REG) (also known, in more traditional thermoacoustic studies, as a stack) consists of a porous material specifically designed to facilitate thermo-acoustic energy conversion. Indeed, from thermoacoustic principles, it is well-known that the energy conversion is particularly significant when in presence of a spatial temperature gradient imposed on the regenerator; a common setup for thermoacoustic engine applications \cite{SwiftBook,Scalo,Chen0,HaoJAP}. In the following we will show that, other than powering TA engines, the TA coupling can also be leveraged to manipulate the propagation of TA Bloch waves and to shape its dispersion. 

One of the most immediate consequences of the periodic exchange of thermoacoustic energy and of the isothermal condition imposed to the regenerators, the periodic TA waveguide is non-conservative and can either result in an effective lossy or amplifying medium (depending on the direction of propagation of the wave with respect to the thermal gradient).
While every natural material includes some degree of non-conservative effects (the most immediate being the mechanical energy dissipation associated with structural damping), these effects were often deemed negligible so that the early literature on periodic media and metamaterials had predominantly focused on the analysis of lossless conservative media. In this context, the classical dispersion analysis led to a real-valued band structure (RBS) for all propagating modes. 
In recent years, however, the intentional use of non-conservative effects has started drawing considerable attention and has established itself as a possible way to further manipulate the wave propagation characteristics of the host medium . Typical examples consist in metamaterial systems exploiting viscoelastic inserts as one of the constitutive phases \cite{Wang15, Liu, Oh}, or in dissipative periodic acoustic waveguides \cite{Theocharis,Henriquez}. For this class of lossy materials, the RBS approach was not applicable anymore, and the analysis of the dispersion behavior required a complex-valued band structure (CBS) approach \cite{Wang15}. In CBS, the real part of the solution characterizes the propagating waves, while the imaginary part captures either the dissipation or the spatial attenuation of the waves.

As previously mentioned, in TA systems a significant energy exchange occurs between the mechanical component (carried by the acoustic wave) and the thermal component (produced by the heat source), hence giving rise to non-conservative behavior. It follows that, as for the example of viscoelastic metamaterials \cite{Wang15}, the analysis of the band structure in TA periodic systems will also require a CBS approach. In addition, and differently from viscoleastic metamaterials, TA periodic systems are also intrinsically non-reciprocal (due to the presence of the spatial thermal gradient on the REG). We will show that CBS is still well equipped to capture the response of TA periodic system and that its application allows uncovering the existence of an anomalous unidirectional energy transport phenomenon. The dispersion calculation will be followed by an analysis, in the long-wavelength limit, of the effective properties of the TA unit cell. Results will reveal the ability of the TA metamaterial to act as a single-zero effective medium which, in turns, leads to a zero refraction index material. Interestingly, these effective properties and the bandwidth of effective zero refractive index can be tuned via controlling the intensity of the thermal gradient, and can potentially open a route to a tunable single (and, possibly, double) zero medium to achieve, among others, energy squeezing and cloaking effects.

Finally, but not less significant, the model approach and analysis proposed in this study furthers the understanding of the amplification and attenuation characteristics of TA waves in the small-channel limit, which may have significant implications for the optimal design of thermoacoustic amplifiers \cite{Senga} and diodes \cite{Biwa16} and, more in general, for TA engines in traveling wave configuration. 

\section{Problem Statement \label{PS}}

\begin{figure}
    \centering
    \includegraphics[width=\linewidth]{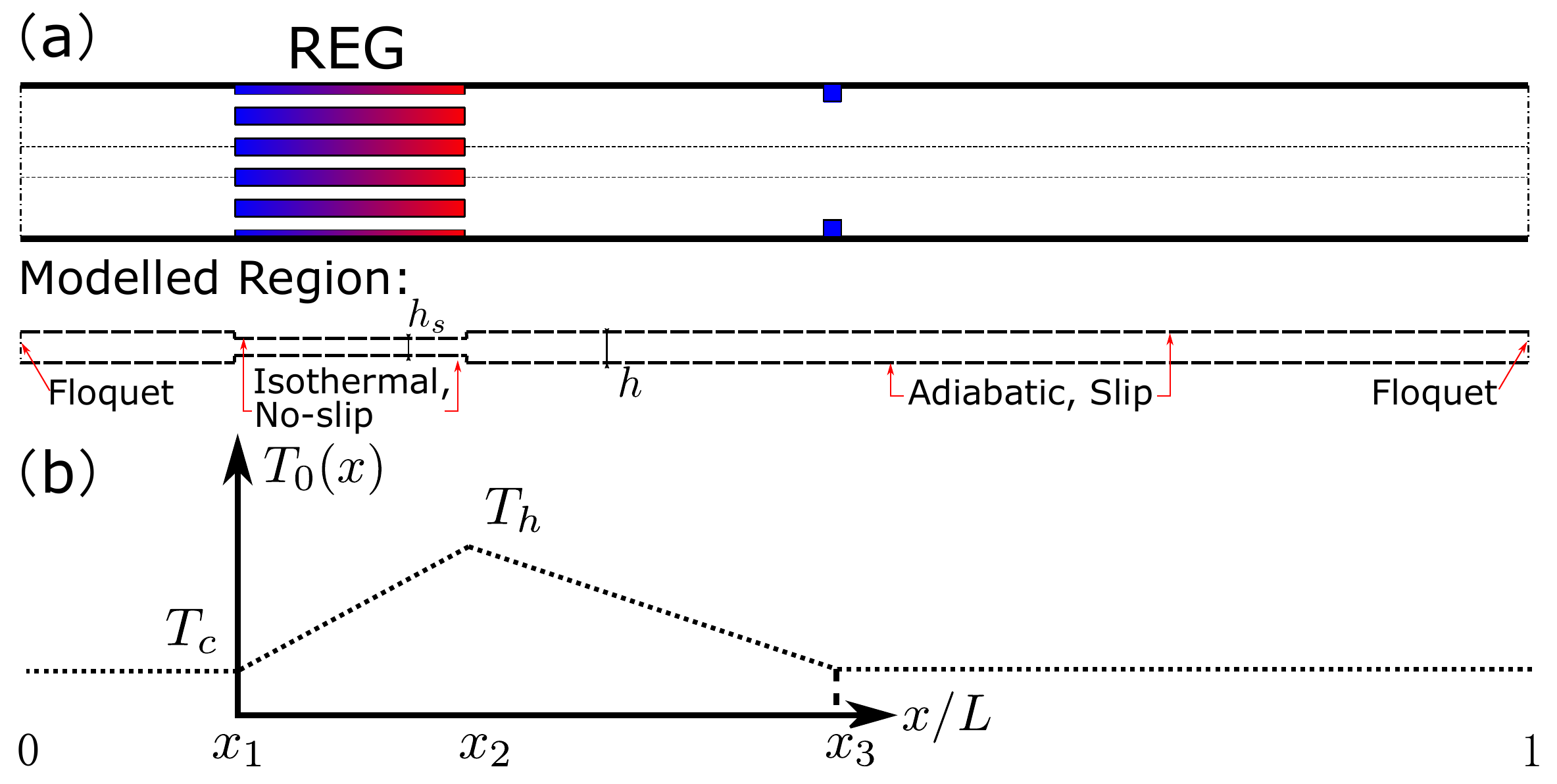}
    \caption{(a) (top) Schematic of the unit cell of the periodic thermoacoustic waveguide, and (bottom) the section of the waveguide being modelled. (b) The mean temperature distribution along the unit cell. A spatial temperature gradient is imposed on the REG. An ambient heat exchanger is located at $x/L=x_3$.}
    \label{schematic}
\end{figure}

The system under investigation consists of a one-dimensional infinite periodic waveguide whose fundamental unit cell is made of a straight duct and a regenerator (see Fig. \ref{schematic}(a)). The regenerator can be thought of as a stack of short parallel plates separated by thin pores so that, at low frequency, viscous and thermal conduction losses cannot be neglected. A temperature spatial gradient is imposed on the regenerator (Fig. \ref{schematic}(b)) so to elevate the temperature from ambient temperature $T_c$ at one end to the hot temperature $T_h$ at the other end. The hot end of the regenerator is followed by a thermal buffer tube (TBT), terminated by an ambient heat exchanger that allows recovering the reference ambient temperature $T_c$. The resulting temperature distribution in the unit cell is plotted in Fig. \ref{schematic}(b). Note that the TBT enables a continuity of temperature at the cell ends, while also acting as a local scatterer due to the temperature variation from $T_c$. 

In order to analyze the system, and without losing generality, we adopt a plane wave assumption for the wave propagating in the 1D duct outside the regenerator. This same assumption is not valid for the waves inside the small pores of the regenerator due to the thermo-viscous effects. Therefore, inside the regenerator channels, the solution is developed according to Rott's thermoacoustic linear theory \cite{Rott,SwiftBook}:
\begin{align}
    \frac{\mathrm{d}p}{\mathrm{d}x}&=-\frac{\rho_0}{1-f_v}(\mathrm{i}\omega)u& \label{mom}\\
    \frac{\mathrm{d}u}{\mathrm{d}x}&=-\frac{1+(\gamma-1)f_k}{\gamma P_0}(\mathrm{i}\omega)p+g u \label{con}
\end{align}
where
\begin{equation}
    g=\frac{f_k-f_v}{(1-f_v)(1-\mathrm{Pr})}\frac{1}{T_0}\frac{\mathrm{d}T_0}{\mathrm{d}x} \label{g}
\end{equation}
$u$ and $p$ are first-order cross-sectionally averaged particle velocity and pressure, respectively. $\rho_0, P_0$ and $T_0$ are zeroth-order (mean-state) density, pressure and temperature, respectively. $\gamma$ and Pr are specific heat ratio and Prandtl number, respectively. $f_k$ and $f_v$ are complex functions expressed as: 
\begin{equation}
    f_{\Box}=\frac{\mathrm{tanh[(1+i)}(h/2)/\delta_{\Box}]}{\mathrm{[(1+i)}(h/2)/\delta_{\Box}]}
\end{equation}
where ${\Box}$ can take either the subscript of $v$ or $k$. $h$ is the width of the straight section. The viscous and thermal penetration depths are expressed as:
\begin{align}
    \delta_v=\sqrt{2\nu/\omega},\quad\quad\delta_k=\sqrt{2\kappa/\omega} \label{delta}
\end{align}
 where $\nu$ and $\kappa$ are the dynamic viscosity and the thermal diffusivity, respectively. Note that the thermo-viscous coupling is particularly strong when the characteristic ratio $h/2\delta_*$ is small. This latter condition can occur either when in presence of thin channels (i.e. small $h$) or of low frequency waves (i.e. large $\delta_*$, Eqn \ref{delta}). When the thermo-viscous effects are negligible, that is when $f_v=f_k=0$, the Helmholtz equation is recovered from Eqns.$~$\ref{mom} and \ref{con}. Considering the plane wave assumption for the wide sections (outside the regenerator), as well as the fact that the pores in the regenerator are identical to each other, we simplify the modeling by only calculating the acoustic field in a minimal unit (including a single pore) \cite{Gupta}, outlined by the dashed lines in Fig. \ref{schematic}. The Floquet boundary conditions are applied to the cell (minimal unit) ends:
\begin{align}
    u(L)&=\mathrm{exp}(-\mathrm{i}kL)u(0)\label{Fl1}\\
    p(L)&=\mathrm{exp}(-\mathrm{i}kL)p(0) \label{Fl2}
\end{align}
Recall that the wavenumber $k$ is complex-valued for the CBS analysis.

Mathematically, the CBS could be formulated in three different ways: (1) complex $\omega$ (frequency) versus real $k$ (wavenumber), (2) real $\omega$ versus complex $k$, and (3) complex $\omega$ versus complex $k$. However, only the former two representations are physically significant. The first representation, appropriate for free wave propagation, considers a complex frequency $\omega=\mathrm{Re}[\omega]+\mathrm{iIm}[\omega]$ under a given real wavenumber $k$, where the imaginary part $\mathrm{Im}[\omega]$ denotes the temporal growth ($\mathrm{Im}[\omega]<0$) or decay ($\mathrm{Im}[\omega]>0$) of the transient wave. The complex frequency is especially relevant to TA engines in order to describe the transient exponentially growing motion due to TA instability  \cite{Penelet,Lin,ChenATE19,HaoJSV20,HaoMSSP}. The imaginary part of $\omega$ is also widely used to represent the decay rate of the free vibration of a lossy material. The second representation is better suited for a time-harmonic wave propagation. The imaginary part of the complex wavenumber $k=\mathrm{Re}[k]+\mathrm{iIm}[k]$ allows capturing either the spatial attenuation or amplification of the time-harmonic wave at steady state. In this study, we adopt this latter (time-harmonic) description of CBS in which a forcing frequency $\omega$ is taken as the real independent variable for the solution of a complex wavenumber $k=\mathrm{Re}[k]+\mathrm{iIm}[k]$ \cite{Wang15,Suzuki}, in order to describe either the spatial amplification or attenuation of thermoacoustic Bloch waves. 

The discretization of Eqns$~$\ref{mom} and \ref{con} (for the REG and for the duct, respectively) combined with the Floquet boundary conditions (Eqns$~$\ref{Fl1} and \ref{Fl2}) yields a generalized eigenvalue problem:
 \begin{equation}
     [\textbf{A}-\mathrm{exp}(-\mathrm{i}kL)\textbf{B}]\begin{bmatrix}
            \textbf{p}\\\textbf{u}
         \end{bmatrix} =0
         \label{GenEigProb}
 \end{equation}
 where \textbf{A} and \textbf{B} are coefficient matrices in which certain elements are frequency-related. The eigenfunction $\textbf{[p, u]}^\mathrm{T}$ consists of the discrete distribution of pressure and velocity. Given the frequency $\omega$, the eigenvalue $\mathrm{exp}(-\mathrm{i}kL)$ can be obtained by applying any available eigenvalue solver. The complex wavenumber $k$ can then be easily extracted. Repeating the process for different values of the frequency $\omega$ spanning a given range leads to the CBS of the system.
 
 Recall that the thermoacoustic coupling results from the combined effect of the thermo-viscous behavior ($f_v\neq0$ and $f_k\neq0$) and of the temperature gradient along the regenerator ($T_h \neq T_c$). For a better understanding of the TA Bloch waves, we perform a CBS analysis of the waveguide under three configurations employing different assumptions: (1) pure acoustics, or equivalently, lossless acoustics ($f_v=f_k=0$, $T_h=T_c$), (2) thermo-viscous acoustics ($f_v\neq0,$~$ f_k\neq0$, $T_h=T_c$), and (3) thermoacoustics ($f_v\neq0,$~$ f_k\neq0$, $T_h=1.5T_c$). In all three cases, both the geometrical and the material properties are maintained the same, as listed in Table \ref{para}. Cases (1) and (2) will serve as a reference to better understand the behavior observed in Case (3) that represents the actual TA Bloch waves. Note that cases (1) and (2) are expected to be reciprocal due to the lack of the thermal gradient. In addition, in Case (1) the imaginary part shall be non-zero only in the band gaps (it will be zero outside a band gap due to the lossless assumption).

\begin{table}[h!]
    \centering
    \begin{tabular}{ccccccc}
         $L\mathrm{[m]}$&$h_s\mathrm{[mm]}$ &$h_s/h$&$x_1$&$x_2$&$x_3$&$P_0\mathrm{[Pa]}$\\\hline
         0.5&0.25&0.75&0.1&0.12&0.45&101325 \\\hline\hline
         $T_c=T_\mathrm{ref}\mathrm{[K]}$  &$\rho_\mathrm{ref}\mathrm{[kg/m^3]}$ & Pr & $\gamma$&\multicolumn{3}{c}{$\mu(T_0)\mathrm{[Pa\cdot s]}$}\\\hline
         300&1.2 & 0.72 & 1.4&\multicolumn{3}{c}{$1.98\times 10^{-5}(T_0/T_{\mathrm{ref}})^{0.76}$}
    \end{tabular}
    \caption{Geometrical and material parameters of the TA unit cell.}
    \label{para}
\end{table}

\section{Results and Discussion}
\begin{figure*}[]
     \centering
     \includegraphics[width=\linewidth]{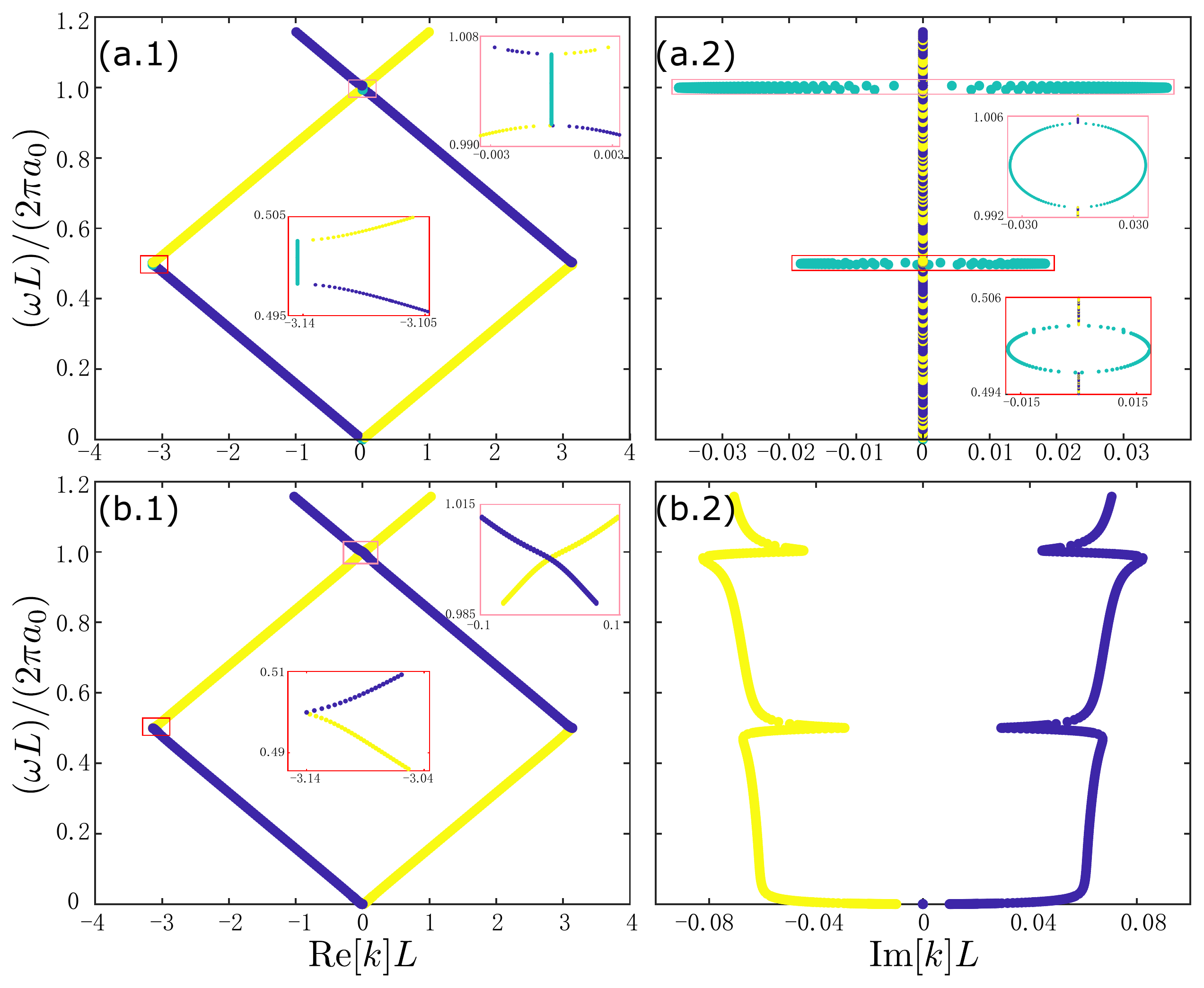}
     \caption{Complex band structure of the unit cell under (a) pure acoustics ($f_k=0,~f_v=0~T_h=T_c$) assumption, and (b) thermo-viscous acoustics ($f_k\neq0,~f_v\neq0~T_h=T_c$) assumption. Reciprocity is preserved under both assumptions. The vertical axis is the reduced frequency, where $a_0$ is the ambient sound speed at room temperature $T_c$, i.e. $a_0=\sqrt{\gamma P_0/\rho_{\mathrm{ref}}}=343 \mathrm{[m/s]}$.}
     \label{PATVA}
 \end{figure*}

Figure \ref{PATVA}(a) and (b) show the CBS of the periodic system under pure acoustics and thermo-viscous acoustics assumptions. Band gaps appear in Fig. \ref{PATVA} (a.1) at the band crossings. These band gaps are the result of Bragg scattering occurring at the abrupt cross-sectional area changes at the REG ends. The scattering is particularly strong when the length of the unit cell is approximately a multiple of the half wavelength. The Im$[k]$ is non-zero in the band gaps, indicating the presence of evanescent waves. The symmetry of the CBS also suggests that reciprocity is preserved. This behavior is clearly not surprising and consistent with the well-known response of classical non-resonant periodic media with periodic mechanical impedance mismatch.
In the CBS plots, modes are colored by the sign (or, equivalently, the direction of propagation) of the cycle-averaged acoustic intensity. The intensity is expressed as $I=0.5\mathrm{Re}[p\overline{u}]$, where the over-bar denotes complex conjugate quantities. Remember that the sign of the acoustic intensity indicates the direction of acoustic energy transport. The colors yellow and blue represent positive and negative intensity, while the green color denotes a zero intensity (which only appears for evanescent modes). The band gaps in Fig. \ref{PATVA} (a.1) disappear in Fig. \ref{PATVA} (b.1) due to the significant thermo-viscous losses in the REG pores. The symmetrical distribution of $k$ is still preserved in these configurations (Fig. \ref{PATVA} (b.1) and (b.2)) because the temperature gradient is not activated. It is worth mentioning that the viscous and heat-conduction losses break time-reversal symmetry, but do not affect reciprocity. It is also notable that in Fig. \ref{PATVA} (b.2) all non-zero Im$[k]$ correspond to spatial attenuation. Considering the $\mathrm{exp}(-\mathrm{i}kL)$ notation for Floquet conditions, a forward propagating wave (with positive intensity, yellow) attenuates along $x$ if $\mathrm{Im}[k]<0$, while a backward propagating wave (with negative intensity, blue) attenuates along $-x$ if $\mathrm{Im}[k]>0$.

\begin{figure*}
     \centering
     \includegraphics[width=\linewidth]{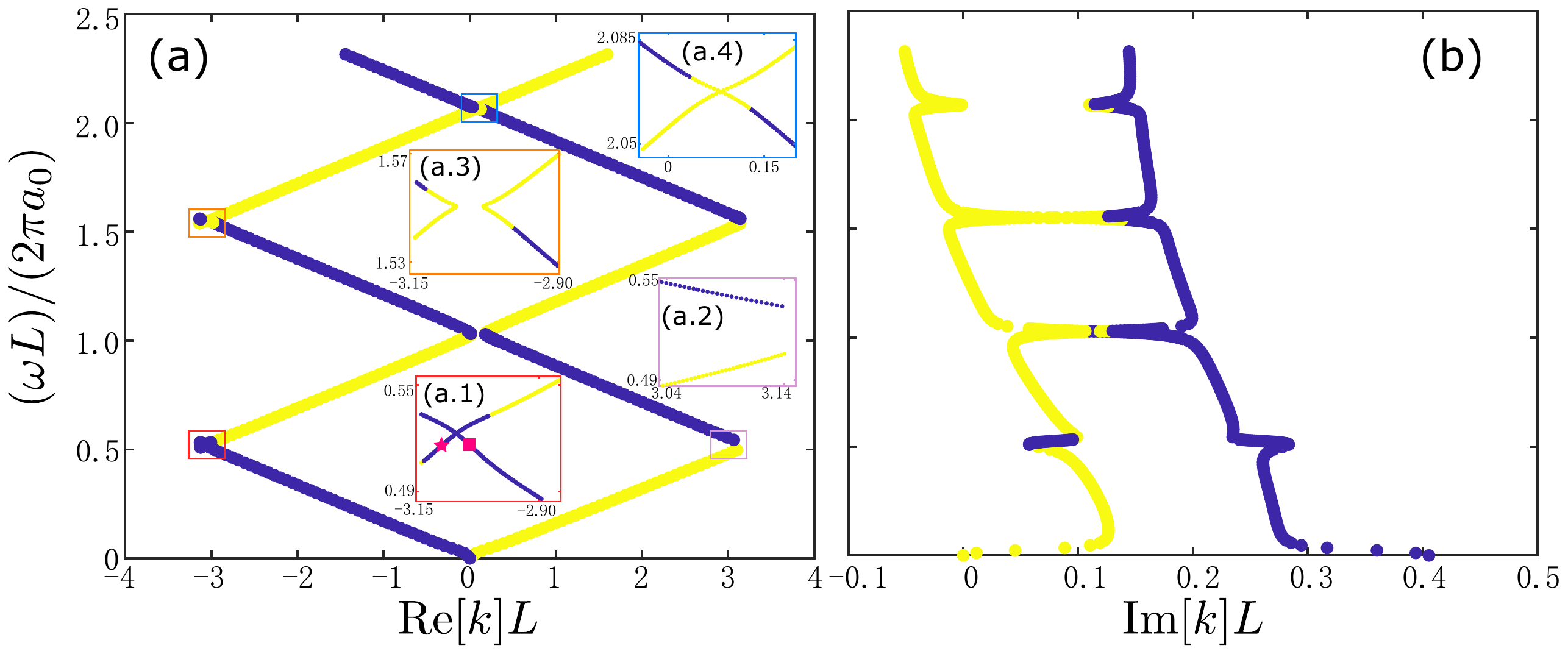}
     \caption{Complex band structure of the unit cell under thermoacoustic ($f_k\neq0,~f_v\neq0~T_h=1.5T_c$) assumption. Reciprocity is broken indicated by the asymmetric CBS. The star and the square in (a.1) denote two modes at the same frequency with opposite direction of group velocity, yet same direction of intensity. The acoustic impedance of these two modes will be plotted in Fig. \ref{impedance}.}
     \label{TACBS}
 \end{figure*}

Figures \ref{TACBS} (a) and (b) show the results for Case 3 in the form of the real and imaginary parts of the CBS (indicated, in the following, as Re[CBS] and Im[CBS]) of the TA Bloch waves under thermoacoustic coupling ($f_v\neq0,$~$ f_k\neq0$, $T_h=1.5T_c$). In the Re[CBS], the band crossing that in the (reciprocal) thermo-viscous case (Fig. \ref{PATVA}(b.1)) occurred at $\mathrm{Re}[kL]=\pm\pi $ is now shifted to $\mathrm{Re}[kL]=-3.07$; as a consequence, an opening appears around $\pi$ in the symmetric half of the first Brillouin zone (BZ) (See Fig. \ref{TACBS}(a.2)). 
The shift of the band crossing is a direct result of space-inversion symmetry breaking, leading to non-reciprocal wave propagation. 

\begin{figure*}
     \centering
     \includegraphics[width=\linewidth]{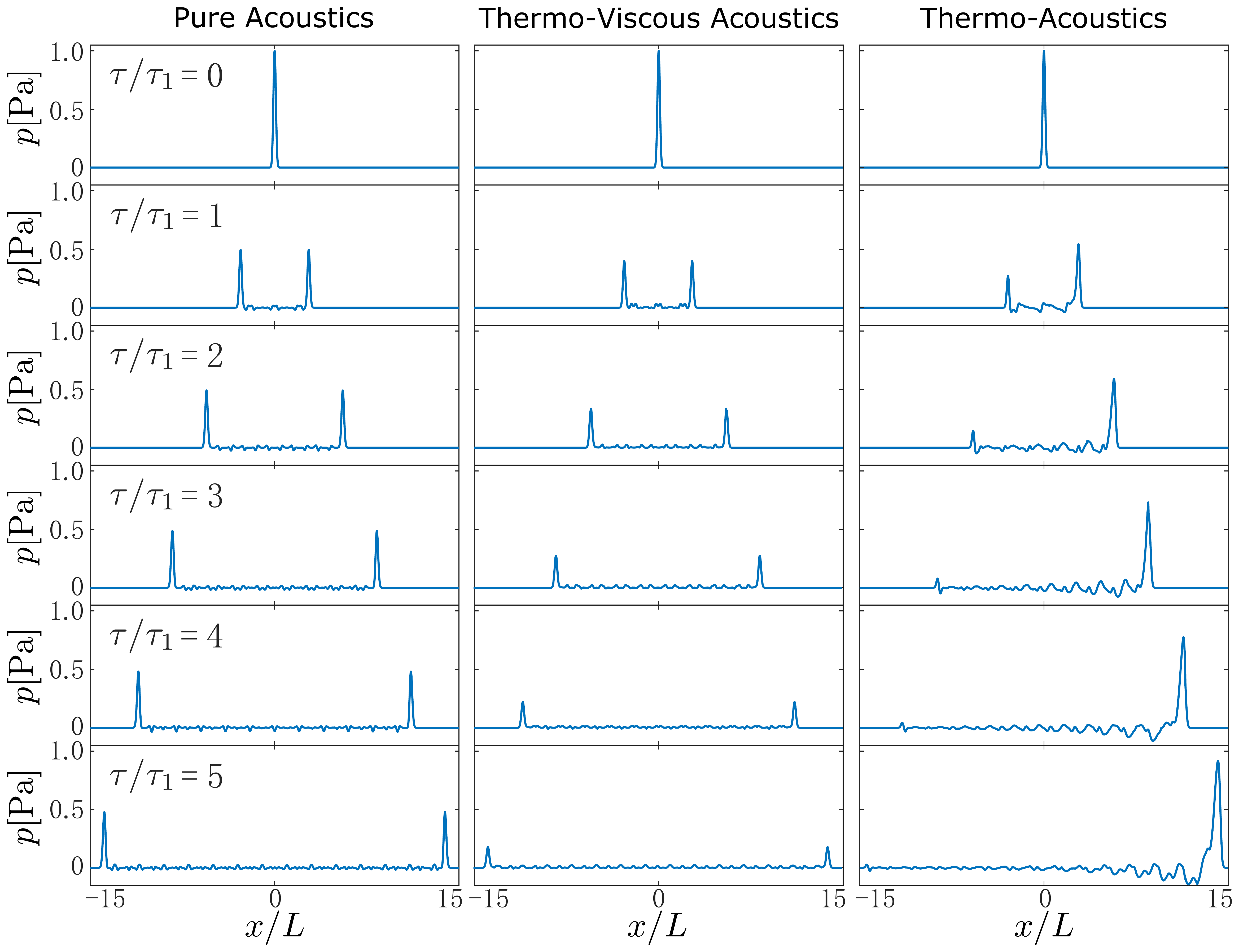}
     \caption{A pressure pulse is applied in the center of a 30-unit TA periodic waveguide. The pulse splits into two fronts propagating in opposite directions. The propagation characteristics are studied for the three different cases previously defined, that is Pure Acoustic (PA), Thermo-Viscous-Acoustic (TVA) and Thermo-Acoustic (TA) assumptions, respectively. In the PA and TVA cases, the two fronts propagate in opposite directions and the response remains symmetric. The TVA response shows amplitude attenuation due to the thermo-viscous losses. In the TA case, the forward propagating front is amplified while the backward propagating one is attenuated due to the TA coupling. The combined effect results in an effective non-reciprocal, unidirectional sound propagation. The small fluctuations near $p=0$ are due to the inter-cell reflections. $\tau$ indicates time and $\tau_1=0.004$[s].}
     \label{pulse}
 \end{figure*}

 Recall that, under the thermo-viscous assumptions, either a forward or a backward propagating wave (marked in yellow or blue in Fig. \ref{PATVA}(b), respectively) is always associated with either a negative or a positive Im[$k$], which indicates spatial attenuation along the direction of propagation of the wave. Unlike the thermo-viscous case, in the TA case the forward propagating waves can be spatially amplified if $T_h\neq T_c$. This latter condition is indicated by yellow dots with positive Im[$k$] in Fig. \ref{TACBS}(b). At the same time, the backward propagating wave still attenuates along $-x$, according to the positive Im[$k$]. Such asymmetric non-conservative behavior (characterized by either attenuation or amplification) is a distinctive feature of the non-reciprocal response of TA Bloch waves. We highlight that the non-reciprocal propagation can be effectively exploited to achieve an effective one-way sound transmission in the 1D TA waveguide. Figure \ref{pulse} shows the simulated results of a pressure pulse applied at the center of a finite length waveguides consisting of 30 unit cells. The propagation of the pulse is studied for the three configurations discussed above, that is Pure Acoustics (PA), Thermo-Viscous Acoustics (TVA), and TA assumptions described in Section \ref{PS}. Clearly, in the PA and TVA cases, the pulse splits in two equal parts that propagate in opposite directions maintaining a symmetric behavior. The same situation occurs for the TVA configuration, however the two pulses experience a spatial attenuation due to thermoviscous losses. The small fluctuations near $p=0$ are due to the inter-cell reflections. In the TA case, after the initial pulse separates into two fronts, the forward propagating front is spatially amplified while the backward moving one is attenuated. The spatial amplification of acoustic waves relies on the heat provided to the system at the hot side of REG. The amplitude of the amplified pulse will eventually reach a steady state value, balanced by nonlinear saturation \cite{Gupta, Tan}. It follows that, in the far field, only the forward propagating pulse survives while the backward one disappears. This one-directional, non-decaying propagation may find interesting applications in long range acoustic communication \cite{Shi}. The spatial amplification of time-harmonic waves in a waveguide formed by a cascade of several unit cells had been experimentally demonstrated in Refs.$~$\onlinecite{Biwa16} and \onlinecite{Senga}, although without providing the underlying theoretical framework. We merely note that the ability of the TA propagating Block wave to reach a consistently stable amplitude could effectively result, in the far field, in a cloaking behavior. Indeed, any reduction in the wave amplitude due to scattering effects would be recovered, upon propagation, due to the TA coupling; hence not leaving any detectable trace of scattering in the far field.
 
 \subsubsection{\label{UET}Unidirectional energy transport}
An inspection of the acoustic intensity associated with the TA coupled modes in the reduced frequency range $(\omega L)/(2\pi a_0)\in[0.51, 0.53]$ reveals that the intensity $I$ of both modes has the same sign (Fig. \ref{TACBS}(a.1)). Recall that $a_0$ is the ambient sound speed in air, i.e. $a_0=\sqrt{\gamma P_0/\rho_{\mathrm{ref}}}=343 \mathrm{[m/s]}$.
In other words, the mode labelled with a star marker has positive group velocity, yet negative intensity. This observation is certainly counter-intuitive because in classical wave theory the group velocity is also understood as the direction of energy transport. The results suggest that, while a wave packet with the carrier frequency around $(\omega L)/(2\pi a_0)=0.52$ propagates forward as a whole, the energy of the carrier frequency time-harmonic TA Bloch wave is transmitted along the direction of the intensity, that is the $-x$ direction. Similar phenomena are also observed around the reduced frequency 1.55 and 2.07 (Fig. \ref{TACBS}(a.3) and (a.4)), where both modes have positive intensity. The inconsistency between the group velocity $v_g$, also dubbed macroscopic energy transport velocity \cite{BradleyTR}, and the microscopic energy transport velocity $v_E$ associated with the acoustic intensity $I$ in acoustic Bloch waves was examined by Bradley \cite{BradleyTR}. Noting that $v_E$ is proportional to $I$. For inviscid acoustic Bloch waves, the sign of $v_E$ and $v_g$ can differ due to the portion of energy stored in scattering elements eventually present on the wave path. In other terms, while the majority of the energy is transported by the wave in the same direction of propagation, a small portion of energy is stored in the wave scatterers. This part of the energy, termed \textit{stagnant energy}, is not accounted for in the calculation of the microscopic energy transport velocity $v_E$. Indeed, the calculation of $v_E$ only considers the portion of energy that is effectively in transport following the time-harmonic wave.  On the other side, the stagnant energy is considered in the calculation of $v_g$, which is a measure of the propagating speed of pulses or wave packets, as well as the energy carried by them.
When the thermoacoustic effect is taken into account, the discrepancy between $v_g$ and $v_E$ is affected by the thermo-viscous effect as well as the thermoacoustic energy production. Although it is the latter effect which mainly contributes to the occurrence of opposite signs for $v_g$ and $v_E$ (or equivalently $I$). 

\begin{figure*}
    \centering
    \includegraphics[width=\linewidth]{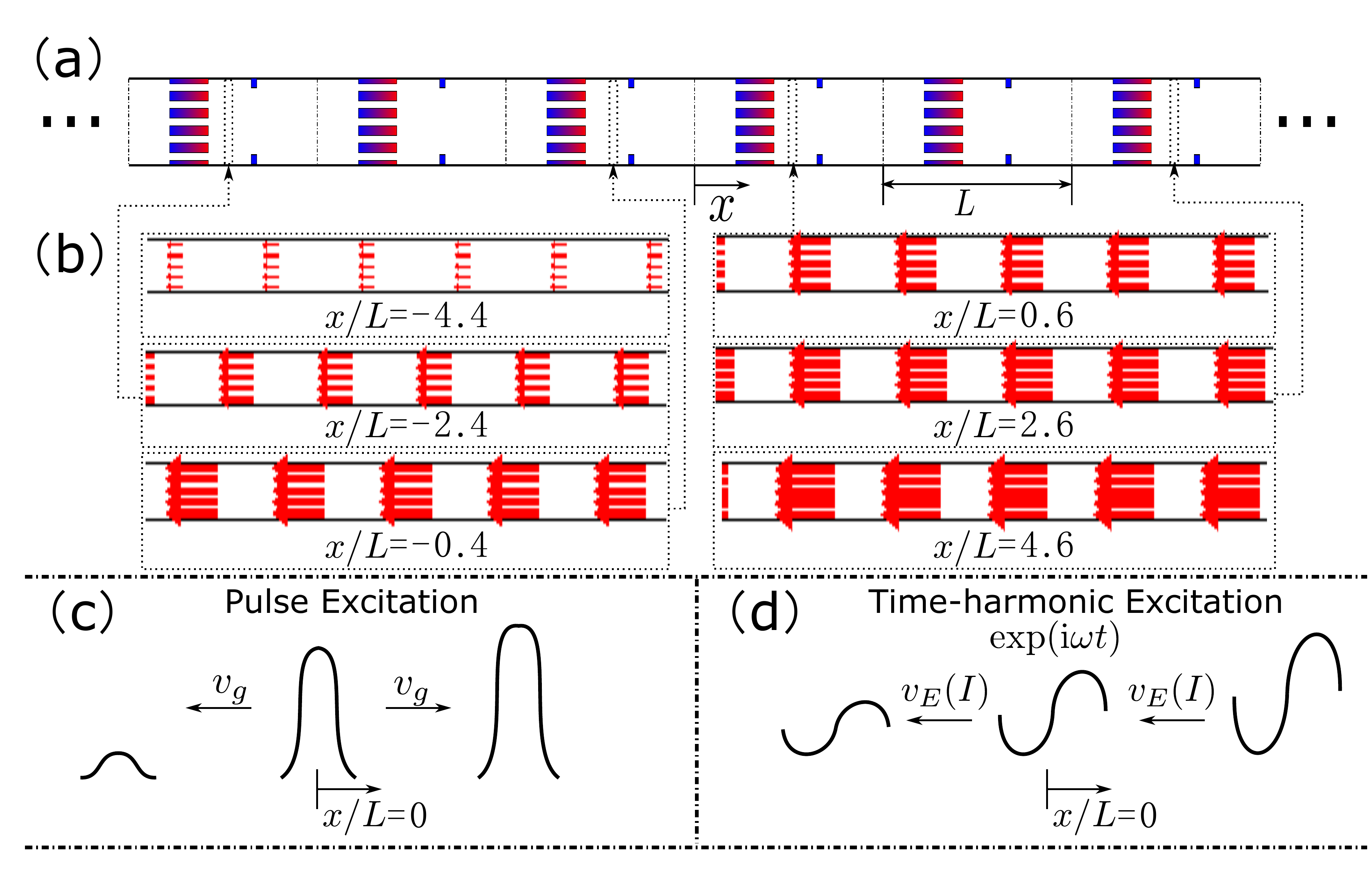}
    \caption{(a) Schematic of a periodic TA waveguide including the REG and the ambient heat exchanger. The color gradient in the REG indicates the cold (blue) and hot (red) temperature side. The vertical dashed lines represent the separation between consecutive unit cells. The behavior of the waveguide is explored under either (c) pulse excitation or (d) time-harmonic excitation. (c) A pulse is generated at $x/L=0$ and splits into two fronts. The two fronts, as well as the energy they carry, travel in opposite directions. The front traveling in the direction of the temperature rise ($x/L>0$ region) is amplified, while the front moving in the opposite direction is attenuated, as shown in Fig. \ref{pulse}. (d) A time harmonic excitation is generated at $x/L=0$ and propagates in both directions. However, at selected frequencies (e.g. the one represented by the star and square markers in Fig. \ref{TACBS} (a.1)) both steady-state modes can transport energy unidirectionally towards $-x$. (b) The cycle-averaged acoustic intensity $I$ associated with the time-harmonic wave is plotted at different locations along the periodic waveguide (highlighted by dashed boxes in (a)). The direction and length of the arrows indicate the direction and log-scale amplitude of the acoustic intensity.}
    \label{energy}
\end{figure*}

To further illustrate the characteristics of the energy transported by time-harmonic waves (at velocity $v_E$) and by a wave packet (at velocity $v_g$), we consider a periodic TA waveguide as shown in Fig. \ref{energy}(a). A pulse initiated at $x/L=0$ (and intrinsically composed of many harmonics) will split into two fronts and travel in both directions, with one front being amplified and the other being attenuated, as shown in Fig. \ref{pulse}. The energy carried by each front travels at a speed $v_g$ (the slope of the curves in Fig. \ref{TACBS}(a)) associated with these two fronts as schematically shown in Fig. \ref{energy}(b). If a time-harmonic excitation is instead generated at $x/L=0$ and at a selected frequency (represented by either the star or the square markers shown in Fig. \ref{TACBS}(a.1)), the steady state energy transfer shows different properties. Although the amplitude of the time-harmonic wave increases in the direction of the rising temperature gradient, or, equivalently, decreases in the opposite direction, the acoustic intensity always has a negative sign. In other terms, the energy always flows in the negative $x$ direction. In order to provide further validation and insights in this unexpected result, we perform both fully numerical analyses as well as theoretical investigations.
Starting with the numerical analyses, we developed a finite element model (FEM) (using the commercial software Comsol Multiphysics) of a TA coupled periodic duct consisting of ten unit cells. An impedance type boundary condition was applied at both ends of the periodic assembly to reduce unwanted reflections. The impedance values were calculated using our theoretical model (Eqn. \ref{GenEigProb}) for the specific modes selected (indicated by the star and square markers in Fig. \ref{TACBS}(a.1)). A unit amplitude time-harmonic velocity excitation was applied at $x/L=0$.
Figure \ref{energy}(d) shows the acoustic intensity extracted from the data produced by the finite element model. The arrows direction and size (plotted in log scale) indicate the direction and magnitude of the intensity, respectively, at different locations along the periodic waveguide. It follows that the wave field on the right and left hand sides of the excitation correspond to the two modes indicated by the star and square markers in Fig. \ref{TACBS}, respectively. Figure \ref{energy}(d) shows that, on both sides of the time-harmonic excitation, the cycle-averaged intensities are negative. On the positive half region of the $x$ axis ($x/L>0$), the energy flows towards the mechanical source placed at $x/L=0$ or, equivalently, the mechanical excitation behaves as an effective energy sink. This counter-intuitive behavior is a result of the rectifying effect produced by the periodic distribution of regenerators (REGs) in the waveguide. In fact, the REGs behave as external heat sources that provide (or extract) energy to (from) the waveguide, hence altering the net transport of acoustic energy and its direction. The intensity plots in Fig. \ref{energy}(d), reveal that both modes are attenuated along the energy flow. Recall that the arrows length is plotted in log scale. By inspecting the wave in the $x/L>0$ range, it can be further concluded that for a time-harmonic TA Bloch wave at steady state, the amplitude in the far field being larger than the amplitude of the mechanical excitation does not guarantee that the TA Bloch wave is spatially amplified from the excitation to the far field. Indeed, in the $x/L>0$ range, the time-harmonic wave is attenuated from the far field towards the excitation, along the energy flow. 

The above considerations made on the ground of numerical results were also corroborated by a rigorous derivation of both $v_g$ and $I$. The derivation is provided here below.  

Evaluating $\mathrm{d}[\overline{p}'u+p\overline{u}']/\mathrm{d}x$ yields:
\begin{widetext}

\begin{align}
    \frac{\mathrm{d}}{\mathrm{d}x}[\overline{p}'u+p\overline{u}']&=E_\text{ta}\\&=\frac{\mathrm{i}\rho_0}{1-\overline{f_v}}(1+\omega\frac{f_v'}{1-\overline{f_v}})\overline{u}u+\frac{\mathrm{i}}{\gamma P_0}[1+(\gamma-1)\overline{f_k}+\omega(\gamma-1)\overline{f_k}']\overline{p}p \nonumber\\
    &+\frac{2\omega\rho0}{|1-f_v|^2}\mathrm{Im}[f_v]\overline{u'}u+\frac{2\omega}{\gamma P_m}(\gamma-1)\mathrm{Im}[f_k]\overline{p'}p + gu\overline{p}'+\overline{g}\overline{u}'p+\overline{g}'\overline{u}p
    \label{ddx}
\end{align}
\end{widetext}
where $E_\text{ta}$ is the mechanical energy distribution under TA coupling, the over-bar denotes complex conjugate quantities and the prime denotes $\partial/\partial \omega$. Integrating Eqn.$~$\ref{ddx} over the unit cell and incorporating the Floquet boundary conditions (Eqns.$~$\ref{Fl1} and \ref{Fl2}) yield:
\begin{widetext}
\begin{equation}
4\mathrm{i}L\frac{\partial\overline{k}}{\partial \omega} \{\frac{1}{4}[\overline{p}(0)u(0)+\overline{u}(0)p(0)]\}\mathrm{exp}(2\mathrm{Im}[k]L)=-[\overline{p'}(0)u(0)+\overline{u'}(0)p(0)]\mathrm{exp}(2\mathrm{Im}[k]L-1)+\int_0^L E_\text{ta} \mathrm{d}x    \label{vgI}
\end{equation}
\end{widetext}
Recall that:
\begin{align}
    v_g&=\frac{\partial \omega}{\partial \mathrm{Re}[k]}=1/\mathrm{Re}[\frac{\partial\overline{k}}{\partial \omega}]\\
    I&=\mathrm{Re}\{\frac{1}{4}[\overline{p}(0)u(0)+\overline{u}(0)p(0)]\}
\end{align}
For propagating pure acoustic waves (Im$[k]=0$, $f_v=f_k=0$, and $T_h=T_c$), Eqn.$~$\ref{vgI} becomes:
\begin{equation}
    I/v_g=\langle E\rangle=\frac{1}{4L}\int_0^L (\rho_0 |u|^2+\frac{1}{\gamma P_m} |p|^2 ) \mathrm{d}x \label{vgI_pa}  
\end{equation}
where $\langle E\rangle$ denotes the spatially averaged mechanical energy along the unit cell. Equation \ref{vgI_pa} is consistent with the observation that the intensity of a propagating time-harmonic acoustic plane wave is always in the same direction as its group velocity. However, the existence of thermo-viscous losses (nonzero $f_v$ and $f_k$) as well as of the temperature gradient (nonzero $g$) gives rise to the discrepancy between the sign of $v_g$ and $I$ (Eqn.$~$\ref{vgI}). 

\begin{figure}
    \centering
    \includegraphics[width=\linewidth]{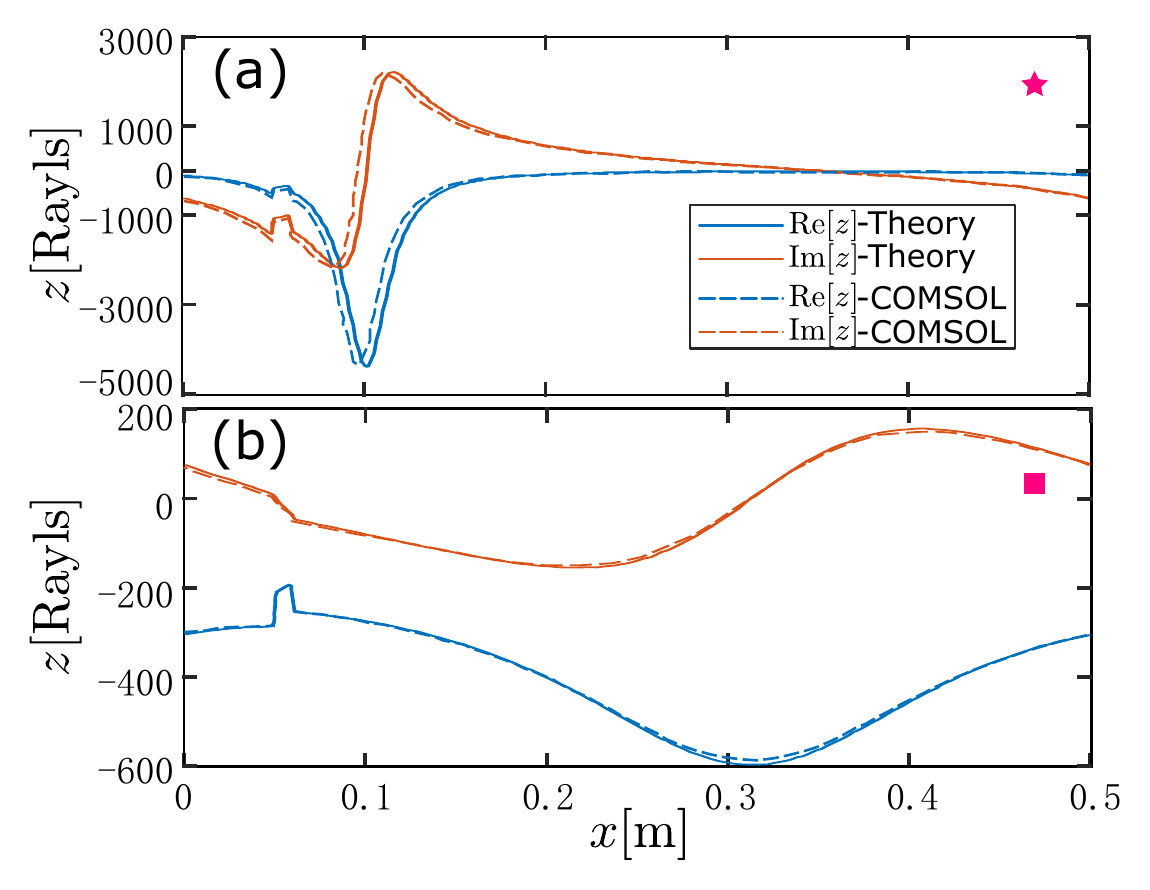}
    \caption{Real and imaginary parts of the impedance $z=p/u$ of the two modes at reduce frequency $(\omega L)/(2\pi a_0)=0.52$, denoted by the star and the square in Fig. \ref{TACBS}(a.1). Both modes show negative Re$[z]$, denoting negative intensity $I$.}
    \label{impedance}
\end{figure}

To further substantiate the previous finding, we plot the acoustic impedance distribution $z=p/u$ (which is scale-independent) along the unit cell for the two modes at $(\omega L)/(2\pi a_0)=0.52$ in Fig. \ref{impedance}. The results are compared with a fully numerical FE solution obtained via COMSOL. The numerical model represents a single pore, as outlined in Fig. \ref{schematic}(a), with parameters given in Table \ref{para}. Similarly to the other numerical FE simulations presented in the earlier part of the section, the right end of the duct was subject to an impedance boundary condition. The complex-valued impedance was extracted from our theoretical results. The duct was excited by applying a unit-amplitude pressure at the left end. The model was used to calculate the steady-state response and to extract the impedance distribution along the waveguide. The results reported in Fig. \ref{impedance} show a very good agreement between the finite element solution and our theoretical calculations. Recalling that $\mathrm{Re}[z]=I/|u|^2$, the negative $\mathrm{Re}[z]$ confirms the occurrence of negative intensity for both modes, hence unidirectional energy transport within this frequency range.

\subsubsection{Effective properties and zero compressibility}
The characterization of TA periodic waveguides also benefits from an analysis of the effective properties.
In the long-wavelength limit, the effective density $\rho_\text{eff}$ and effective compressibility $1/B_\text{eff}$ are expressed as:
\begin{align}
    \rho_\text{eff}&=\mathrm{i}\frac{p(L)-p(0)}{\omega L \langle u \rangle} \label{rhoeff}\\
    1/B_\text{eff}&=\mathrm{i}\frac{u(L)-u(0)}{\omega L \langle p \rangle} \label{compeff}
\end{align}
where the angle brackets denote the spatial average along the unit cell. By comparing Eqn \ref{compeff} and Eqn \ref{con}, we anticipate that the effective compliance can deviate considerably from the reference value $1/\gamma P_0$ due to the existence of the $gu$ term. To enhance the thermoacoustic coupling effect, represented by the $gu$ term, it is preferable to have (1) a sufficiently large temperature gradient, and (2) a locally expanded REG ($h_s/h>1$) to improve the thermo-acoustic energy conversion efficiency of the REG \cite{Jin,deBlok,Senga}. Therefore, in order to clearly illustrate the effect of the TA coupling on the effective properties, we modify the design parameters for the TA unit cell to the following values: $x_1=0.485,~x_2=0.515,~x_3=0.545,~h_s=0.96\mathrm{[mm]}=5.714h,$ and $T_h/T_c=3$. 

\begin{figure*}[!]
    \centering
    \includegraphics[width=\linewidth]{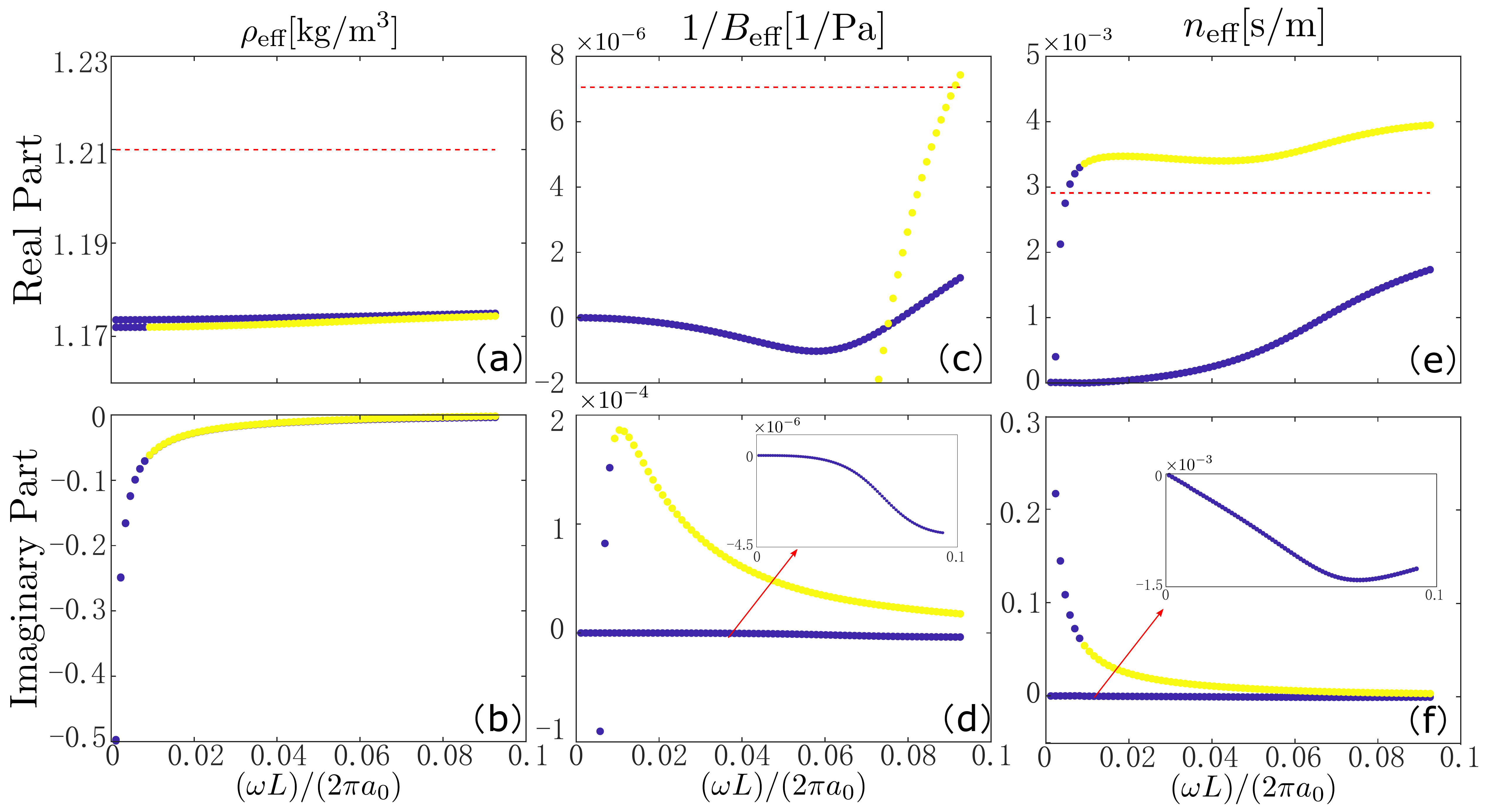}
    \caption{Real and imaginary parts of the effective density $\rho_\text{eff}$, the effective compressibility $1/B_\text{eff}$, and the effective refractive index $n_\text{eff}$. The red dashed lines indicate the reference static values of each property for ambient air at $T_c$.}
    \label{eff_prop}
\end{figure*}

Figure \ref{eff_prop} shows the real and imaginary parts of the effective density, the effective compressibility, and the effective refractive index $n_\text{eff}$ in the TA waveguide. Note that as the frequency decreases, the direction of intensity associated with both modes becomes the same (denoted by the blue color). This phenomenon has been explained in Section \ref{UET}. Due to the non-reciprocity, the effective properties of the two modes are different. The effective density is largely unaffected. The real part is slightly lower than the corresponding static value in ambient air (dashed line) but it remains almost constant for both modes across the selected frequency range. The deviation of Re$[\rho_\text{eff}]$ from the reference value is mainly due to viscous losses ($f_v$ in Eqn. \ref{mom}) and the dependence of the density on the temperature. The imaginary parts of the complex-valued $\rho_\text{eff},~1/B_\text{eff}$ and $n_\text{eff}$, which are a result of the non-conservative thermoacoustic coupling ($f_\Box \neq0$ and $T_h \neq T_c$), affect the spatial amplification or attenuation of the TA Bloch waves. 
The effect of the TA coupling is particularly visible on the effective compressibility. The TA coupling is induced by the term $gu$ in Eqn. \ref{con}, which is proportional to the temperature gradient (Eqn. \ref{g}). By comparing Eqn. \ref{compeff} and Eqn. \ref{con}, it can be seen that the term $gu$ can significantly alter the value of $1/B_\text{eff}$ from the reference value $1/\gamma P_0$.
Figure \ref{eff_prop}(e) shows that the real part of the effective compressibility of one mode (blue) reaches zero in the low frequency range. This result indicates that, due to the TA coupling, the acoustic waveguide behaves as a single zero medium \cite{Dubois, ZhuXF}.
In zero index media, acoustic waves can propagate without phase variation which in turns can lead to acoustic devices exhibiting intriguing properties \cite{Dubois, ZhuXF}, such as acoustic cloaks and energy squeezing \cite{ZhuH}, and acoustic lenses \cite{Kaina}. Double zero properties (i.e. both effective density and effective compressibility) are typically required for an efficient preservation of both phase and amplitude profiles of the acoustic wave front when interacting with scatterers or geometric inhomogeneities. However, single zero media (i.e. zero compressibility) have shown to be capable of preserving the phase profile \cite{Dubois, ZhuXF}. 

The calculation of the refractive index $n_\text{eff}=\sqrt{\rho_\text{eff}/B_\text{eff}}$ shows a near zero Re$[n_\text{eff}]$ value in the frequency range where a near zero Re$[1/B_\text{eff}]$ is achieved. Note that the Im$[n_\text{eff}]$ reflects the spatial amplification or attenuation of the traveling wave traveling but it does not contribute to the phase. 

\begin{figure}
    \centering
    \includegraphics[width=\linewidth]{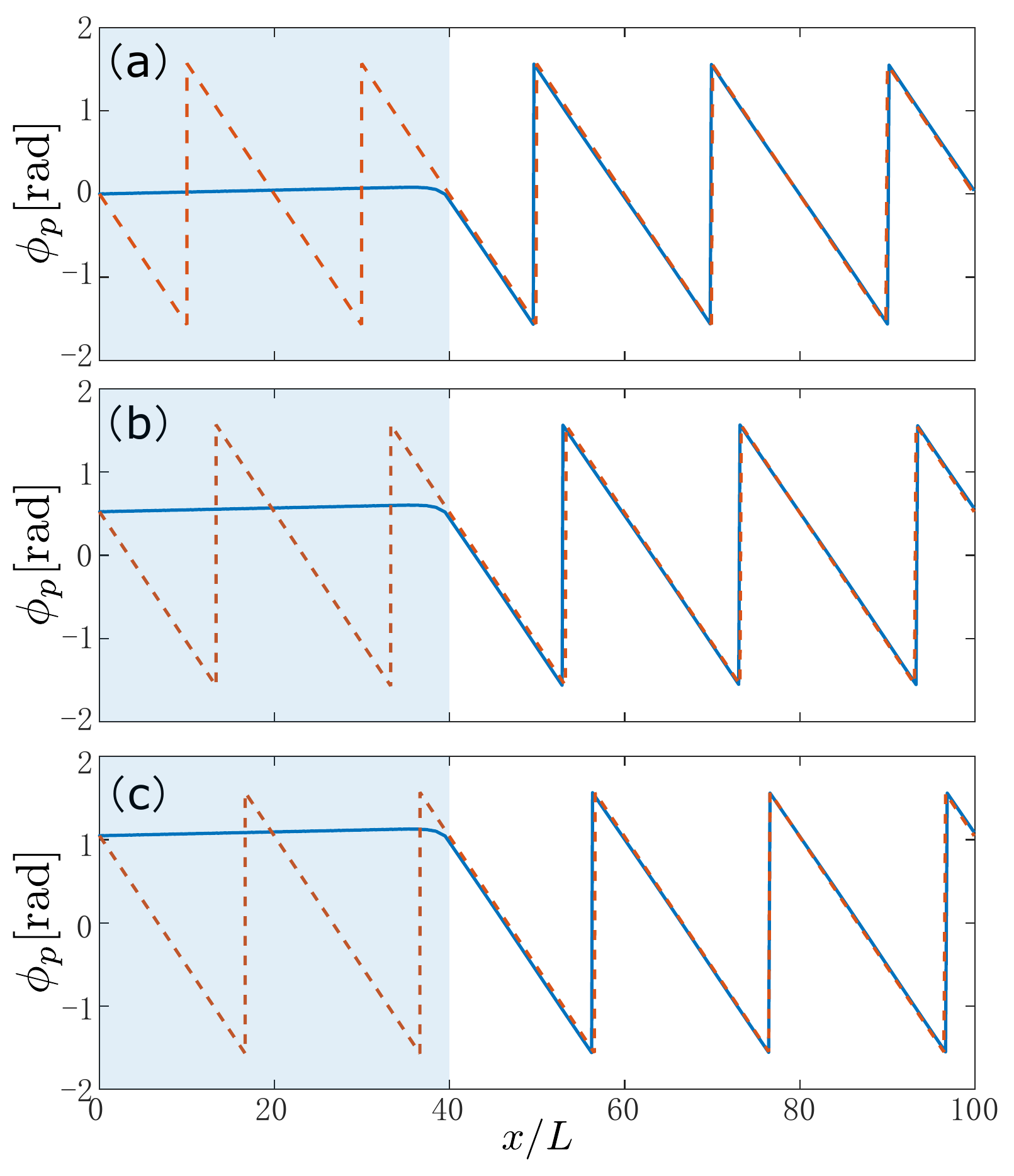}
    \caption{Phase of pressure along the center line of the designed waveguide with TA section (solid), and a hollow duct filled with ambient air as a reference (dashed). The input phase $\theta$ is (a) 0, (b) $\pi/6$ and (c) $\pi/3$. The shaded region represents the TA section.}
    \label{phase}
\end{figure}

In order to further substantiate this observation of the occurrence of a zero-index, we performed a numerical simulation of the steady state harmonic response of the waveguide via finite element modelling. The waveguide consists of 40 TA unit cells ($x/L\in[0,40]$) followed by a homogeneous hollow duct ($x/L\in[40,100]$) filled with ambient air. The hollow duct is $60L$ long and it is terminated at $x/L=100$ with an impedance boundary condition that matches the impedance of air (415 [Rayls]), in order to eliminate reflections. The waveguide is excited at $x/L=0$ with a pressure $p=\mathrm{exp}(\mathrm{i}\theta)$. The solid lines in Fig. \ref{phase} show the phase of the pressure field $\phi_p=\mathrm{atan}(\mathrm{tan}(\mathrm{Im}[p]/\mathrm{Re}[p]))$ along the center line of the waveguide following different input phase $\theta=0,~\pi/6$ and $\pi/3$. The shaded area denotes the section of the waveguide taken by the 40 TA unit cells. The dashed lines show a baseline phase profile that would be obtained if the TA section were substituted by a hollow duct filled with ambient air (that is the entire $100L$ long waveguide would consists in a hollow duct filled with air). Results clearly show that, under near zero refraction index conditions (or, equivalently, near zero compressibility conditions), the phase remains invariant within the TA section, regardless of the input phase.

\begin{figure}
    \centering
    \includegraphics[width=\linewidth]{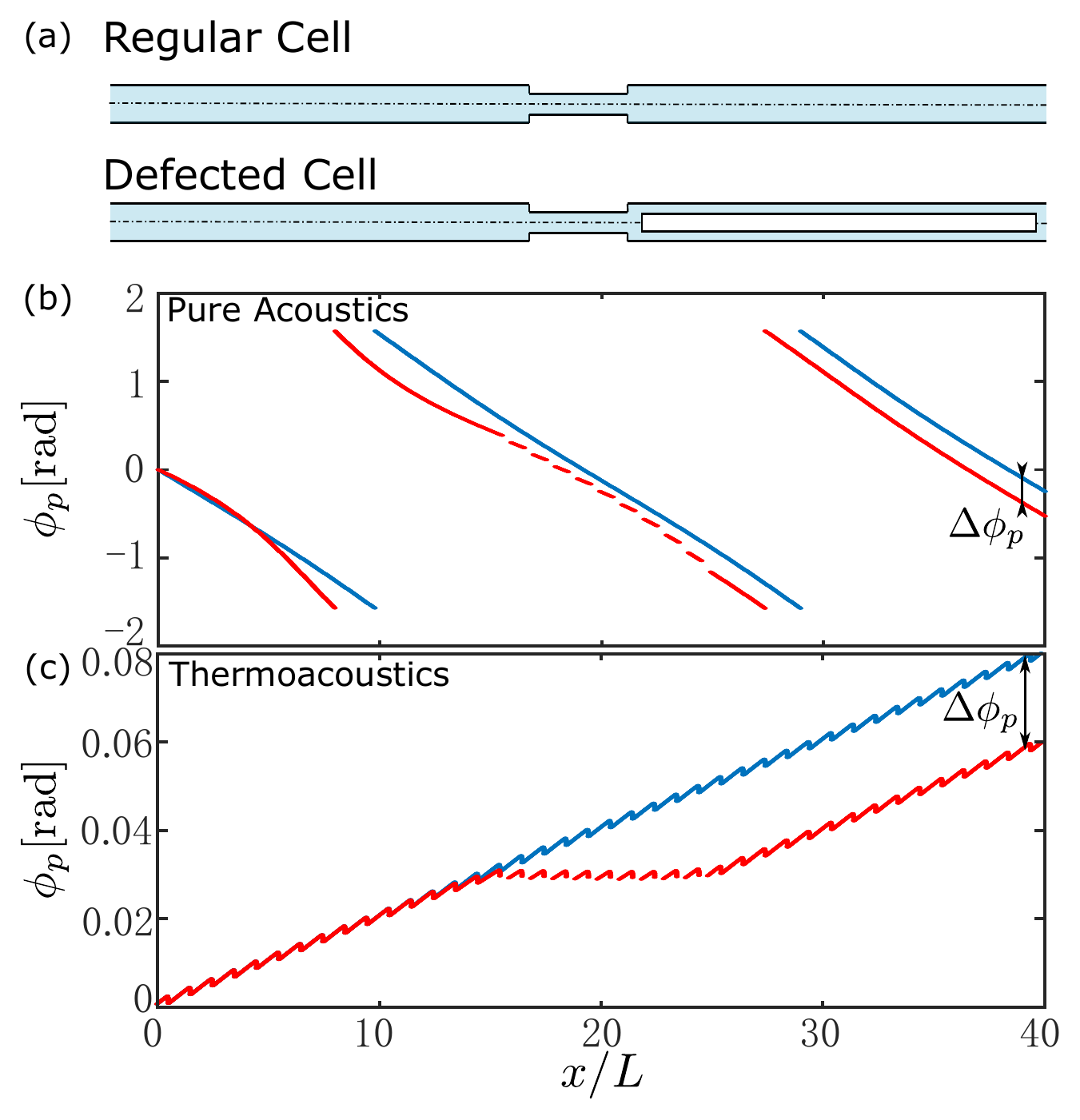}
    \caption{(a) Schematic of a (top) regular unit cell without defect, and (bottom) a defected unit cell with a rectangular sound hard scatterer. (b) and (c) Pressure phase distribution along the centerline (dashed line in (a)) of the 40-cell waveguide under (b) pure acoustic, and (c) thermoacoustic assumptions. The blue and red phase is associated with the regular and the defected waveguide, respectively. The dashed sections in the red curves denote the location of the 10 defected cells.}
    \label{phase_defect}
\end{figure}

An important feature of zero-index materials is that the internal defects do not alter significantly the phase profile \cite{Dubois,ZhuH}. The concept of acoustic cloaking strongly relies on this feature. To further corroborate the phase invariant characteristic of the proposed TA waveguide and its insensitivity to internal defects or scatterers, we performed numerical simulations on an acoustic waveguide composed of 40 unit cells, ten of which included defects. More specifically, 
we consider a baseline configuration made of 40 unit cells without defects (essentially equivalent to the TA waveguide discussed above), and a second configuration in which 10 unit cells in the middle included a defect. The defect consisted in a rectangular sound hard scatterer having dimensions of $0.39L$ long and $0.5h$ wide, as shown in Fig. \ref{phase_defect}(a). The structure was excited at $x/L=0$ with a zero-phase, unit-amplitude pressure, that is $p=\mathrm{exp}(\mathrm{i}0)$. Impedance matching condition is applied to the other end of the structure ($x/L=40$).
The remaining geometrical parameters and the applied temperature gradient are unchanged with respect to the results presented in Figs. \ref{eff_prop} and \ref{phase}.  

Results are presented in Fig. \ref{phase_defect} in terms of phase profile of the pressure field along the centerline (dashed line in Fig. \ref{phase_defect}(a)) of the 40-cell structure. Figure \ref{phase_defect}(b) shows the pure acoustic case (i.e. without TA coupling) which provides a reference baseline. The blue and red curves denote the phase along the regular and the defected waveguide, respectively. From these results it is seen that, in the region downstream of the defects, a 0.26 rad ($\approx15^\circ$) phase difference $\Delta \phi_p$ is induced due to the presence of the defects. Contrarily, when the TA coupling is activated, the shift in phase $\Delta \phi_p$ is reduced to 0.02 rad ($\approx 1^\circ$), as shown in Fig. \ref{phase_defect}(c). This drastic reduction leads to an overall phase shift that is negligible in practical applications. 
We conclude that, as expected based on the previously discussed effective properties, the TA waveguide is still capable of maintaining an effective invariant phase transition, a necessary condition for acoustic cloaking. Recall that the TA waveguide is only a single-zero material, hence incapable of impedance matching, which is a key feature of double-zero materials for high energy transmission \cite{Dubois,ZhuH}. Despite this latter aspect, an interesting phenomenon can take place in this class of waveguides thanks to the multi-physics coupling and to the energy provided to the system by the thermal sources (at the REG locations). Indeed, it was previously shown that the wave traveling in the direction of the rising temperature gradient can be effectively amplified. We also know that, in TA system, sound amplification (due to the underlying TA instability) is balanced by nonlinear losses, ultimately allowing to reach a steady state response \cite{SwiftBook,Gupta}. It is possible to envision that, under this scenario, the amplitude decrease in the transmitted wave (due to the scattering element) can be recovered due to the TA coupling. This means that, in the far field downstream of the scattering section, the response of the two waveguides (i.e. the baseline and the TA configurations) would be exactly identical (both in terms of amplitude and phase profile), hence resulting in a perfect cloaking of the upstream scatterers. In other terms, the lack of impedance matching due to the presence of double-zero effective properties is balanced by the TA growth mechanism. Note that, we did not show a numerical validation of these results because our model does not integrate nonlinear losses.

\begin{figure}
    \centering
    \includegraphics[width=\linewidth]{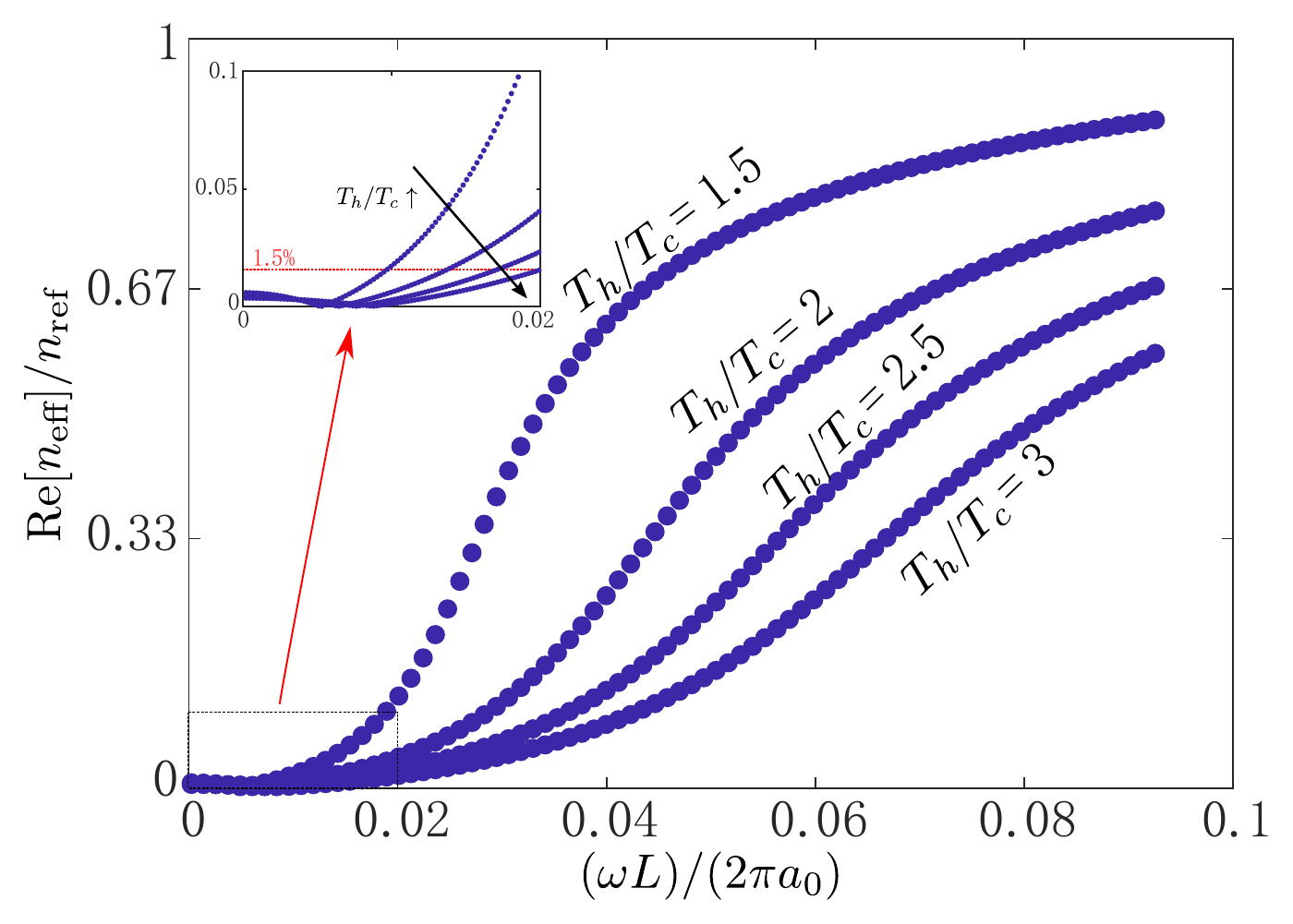}
    \caption{$\mathrm{Re}[n_\text{eff}]$ normalized by $n_\text{ref}$, as a function of the reduced frequency under different temperature gradients, $T_h/T_c$. $n_\text{ref}$ is the refractive index of ambient air at $T_c$, shown as the horizontal dashed line in Fig. \ref{eff_prop}(e). Results show that the effective refractive index of the TA waveguide and the bandwidth of the zero refractive index can be tuned by controlling the temperature ratio $T_h/T_c$.}
    \label{RI}
\end{figure}

An additional interesting aspect of this class of 1D acoustic metamaterials consists in the ability to tune the effective properties by adjusting the intensity of the temperature gradient. Figure \ref{eff_prop}(e) shows that only the lower mode is capable of effective zero refractive index in the low frequency range. Therefore, the tuning effect of the temperature gradient $T_h/T_c$ is illustrated for this specific mode. Using the regular TA waveguide illustrated above, we performed a parametric study by varying the temprature ratio $T_h/T_c$. The numerical results presented in Fig. \ref{RI} show that as the temperature ratio increases, which is indicative of a more intense thermoacoustic coupling, the refractive index $\mathrm{Re}[n_\text{eff}]$ decreases. More remarkably, the effective zero range, that is the range where $\mathrm{Re}[n_\text{eff}]\approx 0$, is expanded as $T_h/T_c$ increases. When $T_h/T_c=3$, the refractive index is effectively zero in the reduced frequency range of approximately $[0\sim0.02]$ where $\mathrm{Re}[n_\text{eff}]$ is less than 1.5\% of $n_\text{eff}$ (Fig. \ref{RI} inset). $n_\text{ref}$ is the refractive index of ambient air at $T_c$, shown as the horizontal dashed line in Fig. \ref{eff_prop}(e). This range spans $\pm 100\%$ of the center (reduced) frequency, 0.01, which is also indicative of the significant potential of TA waveguides to achieve broadband control. Indeed, this latter characteristic stems also from the non-resonant nature of the present design.

We merely suggest that the effective zero compressibility of the TA waveguide could be further combined with effective zero density, realized by resonating side branches \cite{Lee}, to achieve a complete impedance matching double-zero-index material \cite{ZhuH}.

 \subsubsection{Wave amplification and attenuation in small channels}
It is widely accepted in thermoacoustics that, when a temperature gradient is established along a small channel (small-channel limit, such as those in a REG), the volumetric velocity is \textit{proportionally} amplified (attenuated) if the direction of propagation occurs along (against) the positive (i.e. rising) temperature gradient direction \cite{Ceperley,SwiftBook}. This fact can be expressed by the relation $U_h/U_c=T_h/T_c$, where $U$ and $T$ are volumetric velocity and temperature, while the subscripts denote the \textit{h}ot and \textit{c}old ends of the small channel. We anticipate that, thanks to the modeling approach presented in the previous sections, we will be able to make an important discovery concerning the ideal limit behavior of the TA response. Indeed we will show that, unlike the conventional understanding of the small-channel limit behavior in classical thermoacoustics, the \textit{proportional} relation ($U_h/U_c=T_h/T_c$) only holds for the spatially attenuating TA Bloch wave that propagates against the temperature rise. The proportional amplification along the temperature rise does not take place. This very important observation may have great implications for the optimal design of thermoacoustic amplifiers \cite{Senga}.
 
Figure \ref{TACBS}(b) shows that, as $\omega\rightarrow0$, one branch of $\mathrm{Im}[k]L$ converges to a finite value, while the other branch converges to zero. This indicates that one mode propagates with an amplitude variation ($\mathrm{Im}[k]L\neq 0$), while the other mode propagates with an unchanged amplitude ($\mathrm{Im}[k]L=0$). Whether the mode subject to amplitude variation undergoes spatial amplification or attenuation depends on the propagation direction of the mode.
In the following, we theoretically prove that this limit behavior is always true and that the branch of $\mathrm{Im}[k]L$ converging to a finite value, proved to be $\mathrm{Im}[k]L=\mathrm{ln}(T_h/T_c)$, is always associated with a negative intensity, indicative of a spatial attenuation.

As $\omega\rightarrow0$, $h/2\delta_*\rightarrow0$, $f_v\rightarrow[1-(2/3)(h/2)^2(\omega/2\nu)]$ under second order approximation, hence, neglecting the higher order terms, Eqns.$~$\ref{mom} and \ref{con} can be recast into:
\begin{align}
    \frac{\mathrm{d}p}{\mathrm{d}x}&=-\frac{3\rho_0\nu}{(h/2)^2}u \label{mom_0om}\\
    \frac{\mathrm{d}u}{\mathrm{d}x}&=-\frac{\mathrm{i}\omega}{P_m}p+\frac{\mathrm{d}T_m}{T_m\mathrm{d}x}u \label{con_0om}
\end{align}

Considering that the effect of the temperature gradient overpower the viscous effect, Eqn$~$\ref{con_0om} becomes:
\begin{equation}
u_h=(T_h/T_c)u_c
\end{equation}
where $u_h$ and $u_c$ are the cross-sectionally averaged particle velocities at the REG ends.

Two solutions are possible: (1) $u_h=u_c=0$, and (2) $u_h/u_c=T_h/T_c$. Solution (1) is trivial and leads to a constant pressure $p$ distribution according to Eqn.$~$\ref{mom_0om}, or equivalently, $\mathrm{Im}[k]L=0$. Solution (2), under the long-wavelength assumption (small Re$[k]L$), leads to $u(L)/u(0)=u_h/u_c=T_h/T_c=\mathrm{exp}(\mathrm{Im}[k]L)$. Therefore, $\mathrm{Im}[k]L=\mathrm{ln}(T_h/T_c)$. This conclusion is also consistent with the classical understanding of thermoacoustic waves in the "small-channel limit" ($h/2\delta_*\rightarrow0$) \cite{Ceperley}, which in Swift's words is stated as: \textit{"The volume flow rate is amplified in proportion to the temperature rise (or attenuated in proportion to a temperature drop)."} It is understood, based on the previous analysis, that as $\omega\rightarrow 0$ there can be only one wave type (either the forward or backward propagating wave) satisfying $u_h/u_c=T_h/T_c$.

In the following, we show that in the small-channel limit ($\omega\rightarrow 0$) only the ideal proportional attenuation along the temperature drop (associated with a negative intensity, or equivalently, backward propagating wave) is possible. 

For the mode satisfying $u_h/u_c=T_h/T_c$, the velocity distribution in the REG is proportional to $T_0(x)$, i.e.:
\begin{equation}
    u(\xi)=C\big[a\xi+T_c\big]
\end{equation}
where $\xi=x-x_c$, $a=(T_h-T_c)/(x_h-x_c)$, and C is an arbitrary proportional constant. According to Eqn.$~$\ref{mom_0om}, the pressure $p$ in the REG is expressed as:
\begin{equation}
    p(\xi)=-\frac{3\rho_0\nu}{(h/2)^2}C\bigg[\frac{1}{2}a\xi^2+T_c\xi+\frac{(T_h+T_c)Tc}{2a}\bigg]
\end{equation}
Therefore, the intensity at the cold end of the REG ($\xi=0$) is:
\begin{equation}
    I=\frac{1}{2}\mathrm{Re}[p\overline{u}]=-\frac{1}{2}\frac{3\rho_0\nu}{(h/2)^2}|C|^2\frac{(T_h+T_c)T_c^2}{2a}<0 \label{I}
\end{equation}

\begin{figure}
    \centering
    \includegraphics[width=\linewidth]{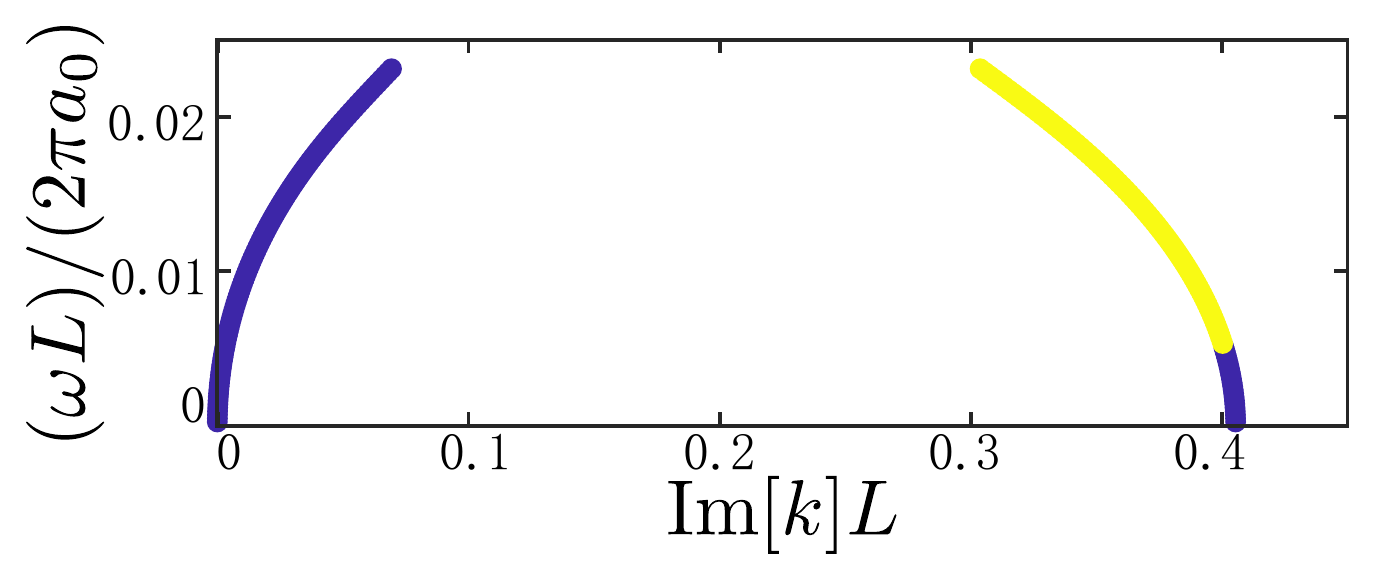}
    \caption{Imaginary part of the CBS for the waveguide in TA configuration under a prescribed set of parameters (see the text). Results show that the ideal amplification of $u$ along the direction of the temperature rise (yellow-colored solution with the value of Im$[k]L=\mathrm{ln}[T_h/T_c]=0.405$ as $\omega \rightarrow 0$) can be closely approached but never reached.}
    \label{TA2}
\end{figure}

Note that, in the case of thermoacoustic amplifiers, it is always desirable to achieve the maximum amplification factor, which according to classical thermoacoustics is $u_\text{out}/u_\text{in}=u_h/u_c=T_h/T_c$ for a single TA unit \cite{Ceperley,SwiftBook}. Based on the conventional understanding \cite{Ceperley,SwiftBook}, this theoretical extreme can be achieved in the small-channel limit. However, the previous TA Bloch wave analysis shows that such ideal change of $u$ proportional to the temperature gradient is always accompanied by a negative intensity $I$ (indicating a spatial attenuation of the wave along $-x$). Nevertheless, by careful design, one can still get close to (although never reach) the ideal amplification of $u$ along the temperature rise. Fig. \ref{TA2} shows the imaginary part of the CBS of a TA Bloch wave in the low-frequency range obtained with another set of parameters: $x_1=0.1,~x_2=0.13,~x_3=0.16,~h_s=0.96\mathrm{[mm]}=5.714h$. Note that the ratio $h_s/h=5.714>1$ represents a locally enlarged REG. The parameters were modified with respect to the anlaysis in Fig. \ref{TACBS} because, according to Ceperley \cite{Ceperley} and Senga and Hasegawa \cite{Senga}, the higher impedance induced by an enlarged REG improves the thermo-acoustic energy conversion efficiency, which is beneficial to the wave amplification in the low-frequency range.
It can be seen that, although the right branch emanating from Im$[k]L$=ln(1.5)=0.405 has negative intensity (blue) as $\omega\rightarrow 0$, the intensity soon becomes positive (yellow) with increasing $\omega$. In other terms, a forward moving Bloch wave, denoted by a positive intensity,  is spatially amplified with a rate that reaches the maximum theoretical amplification rate Im$[k]L\approx$ln($T_h/T_c$). However, as $\omega\rightarrow 0$, the intensity associated with both modes become negative (denoted by blue curves in Fig. \ref{TA2}), which is a sign of unidirectional energy transport. This aspect has been discussed in detail in Section \ref{UET}.

This underlying behavior in the small-channel limit was not observed before and it might have substantial implications to guide the optimal design of thermoacoustic diodes and amplifiers.

\section{Conclusions}
This study presented an in-depth theoretical and numerical  investigation of the dispersion and propagation characteristics of Bloch waves occurring in an acoustic periodic waveguide in presence of thermoacoustic coupling. This class of waves was dubbed thermoacoustic Block waves. The work highlighted several noteworthy findings concerning the basic physical behavior of this wave type as well as their potential impact on future applications. 
While the static temperature gradient imposed on each regenerator led, as expected, to breaking the intrinsic reciprocity of the waveguide, it also highlighted very intriguing and unexpected propagation phenomena. 
Indeed, by leveraging a complex band structure approach, we uncovered an anomalous unidirectional energy transport unique to this type of waveguides. The energy transport was also found to be significantly different depending on the nature of the acoustic wave, that is a wave packet or a harmonic wave, potentially resulting in contrasting directions for the transfer of macroscopic and microscopic energy. 

Also remarkable was the ability of the TA waveguide to act as a broadband, tunable, effective zero refractive index material. In selected frequency ranges, the waveguide was shown to achieve zero effective compressibility, hence zero effective refractive index. In addition, both the effective properties and refractive index could be tuned by simply controlling the intensity of the temperature gradient. The single-zero nature (i.e. zero compressibility) of the TA waveguide was shown to enable phase invariance of the pressure field within the TA waveguide, hence showing potential for application of efficient energy transmission and acoustic cloaking devices. Also very remarkable is the improved understanding of the TA amplification and attenuation effects in the small-channel limit. Specifically, the finding that the amplifying mode can never achieve the theoretical limit could have significant impact on the optimal design of thermoacoustic diodes or amplifiers. 

\bibliography{TAblochWaves}

\begin{thebibliography}{45}%
\makeatletter
\providecommand \@ifxundefined [1]{%
 \@ifx{#1\undefined}
}%
\providecommand \@ifnum [1]{%
 \ifnum #1\expandafter \@firstoftwo
 \else \expandafter \@secondoftwo
 \fi
}%
\providecommand \@ifx [1]{%
 \ifx #1\expandafter \@firstoftwo
 \else \expandafter \@secondoftwo
 \fi
}%
\providecommand \natexlab [1]{#1}%
\providecommand \enquote  [1]{``#1''}%
\providecommand \bibnamefont  [1]{#1}%
\providecommand \bibfnamefont [1]{#1}%
\providecommand \citenamefont [1]{#1}%
\providecommand \href@noop [0]{\@secondoftwo}%
\providecommand \href [0]{\begingroup \@sanitize@url \@href}%
\providecommand \@href[1]{\@@startlink{#1}\@@href}%
\providecommand \@@href[1]{\endgroup#1\@@endlink}%
\providecommand \@sanitize@url [0]{\catcode `\\12\catcode `\$12\catcode
  `\&12\catcode `\#12\catcode `\^12\catcode `\_12\catcode `\%12\relax}%
\providecommand \@@startlink[1]{}%
\providecommand \@@endlink[0]{}%
\providecommand \url  [0]{\begingroup\@sanitize@url \@url }%
\providecommand \@url [1]{\endgroup\@href {#1}{\urlprefix }}%
\providecommand \urlprefix  [0]{URL }%
\providecommand \Eprint [0]{\href }%
\providecommand \doibase [0]{https://doi.org/}%
\providecommand \selectlanguage [0]{\@gobble}%
\providecommand \bibinfo  [0]{\@secondoftwo}%
\providecommand \bibfield  [0]{\@secondoftwo}%
\providecommand \translation [1]{[#1]}%
\providecommand \BibitemOpen [0]{}%
\providecommand \bibitemStop [0]{}%
\providecommand \bibitemNoStop [0]{.\EOS\space}%
\providecommand \EOS [0]{\spacefactor3000\relax}%
\providecommand \BibitemShut  [1]{\csname bibitem#1\endcsname}%
\let\auto@bib@innerbib\@empty
\bibitem [{\citenamefont {Fleury}\ \emph {et~al.}(2014)\citenamefont {Fleury},
  \citenamefont {Sounas}, \citenamefont {Sieck}, \citenamefont {Haberman},\
  and\ \citenamefont {Al{\`u}}}]{Fleury14}%
  \BibitemOpen
  \bibfield  {author} {\bibinfo {author} {\bibfnamefont {R.}~\bibnamefont
  {Fleury}}, \bibinfo {author} {\bibfnamefont {D.~L.}\ \bibnamefont {Sounas}},
  \bibinfo {author} {\bibfnamefont {C.~F.}\ \bibnamefont {Sieck}}, \bibinfo
  {author} {\bibfnamefont {M.~R.}\ \bibnamefont {Haberman}},\ and\ \bibinfo
  {author} {\bibfnamefont {A.}~\bibnamefont {Al{\`u}}},\ }\href@noop {}
  {\bibfield  {journal} {\bibinfo  {journal} {Science}\ }\textbf {\bibinfo
  {volume} {343}},\ \bibinfo {pages} {516} (\bibinfo {year}
  {2014})}\BibitemShut {NoStop}%
\bibitem [{\citenamefont {Fleury}\ \emph {et~al.}(2015)\citenamefont {Fleury},
  \citenamefont {Sounas},\ and\ \citenamefont {Al\`u}}]{Fleury15}%
  \BibitemOpen
  \bibfield  {author} {\bibinfo {author} {\bibfnamefont {R.}~\bibnamefont
  {Fleury}}, \bibinfo {author} {\bibfnamefont {D.~L.}\ \bibnamefont {Sounas}},\
  and\ \bibinfo {author} {\bibfnamefont {A.}~\bibnamefont {Al\`u}},\
  }\href@noop {} {\bibfield  {journal} {\bibinfo  {journal} {Phys. Rev. B}\
  }\textbf {\bibinfo {volume} {91}},\ \bibinfo {pages} {174306} (\bibinfo
  {year} {2015})}\BibitemShut {NoStop}%
\bibitem [{\citenamefont {Swinteck}\ \emph {et~al.}(2015)\citenamefont
  {Swinteck}, \citenamefont {Matsuo}, \citenamefont {Runge}, \citenamefont
  {Vasseur}, \citenamefont {Lucas},\ and\ \citenamefont {Deymier}}]{Swinteck}%
  \BibitemOpen
  \bibfield  {author} {\bibinfo {author} {\bibfnamefont {N.}~\bibnamefont
  {Swinteck}}, \bibinfo {author} {\bibfnamefont {S.}~\bibnamefont {Matsuo}},
  \bibinfo {author} {\bibfnamefont {K.}~\bibnamefont {Runge}}, \bibinfo
  {author} {\bibfnamefont {J.~O.}\ \bibnamefont {Vasseur}}, \bibinfo {author}
  {\bibfnamefont {P.}~\bibnamefont {Lucas}},\ and\ \bibinfo {author}
  {\bibfnamefont {P.~A.}\ \bibnamefont {Deymier}},\ }\href
  {https://doi.org/10.1063/1.4928619} {\bibfield  {journal} {\bibinfo
  {journal} {J. Appl. Phys.}\ }\textbf {\bibinfo {volume} {118}},\ \bibinfo
  {pages} {063103} (\bibinfo {year} {2015})}\BibitemShut {NoStop}%
\bibitem [{\citenamefont {Boechler}\ \emph {et~al.}(2011)\citenamefont
  {Boechler}, \citenamefont {Theocharis},\ and\ \citenamefont
  {Daraio}}]{Boechler}%
  \BibitemOpen
  \bibfield  {author} {\bibinfo {author} {\bibfnamefont {N.}~\bibnamefont
  {Boechler}}, \bibinfo {author} {\bibfnamefont {G.}~\bibnamefont
  {Theocharis}},\ and\ \bibinfo {author} {\bibfnamefont {C.}~\bibnamefont
  {Daraio}},\ }\href@noop {} {\bibfield  {journal} {\bibinfo  {journal}
  {Nature. Mater.}\ }\textbf {\bibinfo {volume} {10}},\ \bibinfo {pages}
  {665–668} (\bibinfo {year} {2011})}\BibitemShut {NoStop}%
\bibitem [{\citenamefont {Liang}\ \emph {et~al.}(2009)\citenamefont {Liang},
  \citenamefont {Yuan},\ and\ \citenamefont {Cheng}}]{Liang09}%
  \BibitemOpen
  \bibfield  {author} {\bibinfo {author} {\bibfnamefont {B.}~\bibnamefont
  {Liang}}, \bibinfo {author} {\bibfnamefont {B.}~\bibnamefont {Yuan}},\ and\
  \bibinfo {author} {\bibfnamefont {J.-c.}\ \bibnamefont {Cheng}},\ }\href@noop
  {} {\bibfield  {journal} {\bibinfo  {journal} {Phys. Rev. Lett.}\ }\textbf
  {\bibinfo {volume} {103}},\ \bibinfo {pages} {104301} (\bibinfo {year}
  {2009})}\BibitemShut {NoStop}%
\bibitem [{\citenamefont {Liang}\ \emph {et~al.}(2010)\citenamefont {Liang},
  \citenamefont {Guo}, \citenamefont {Tu}, \citenamefont {Zhang},\ and\
  \citenamefont {Cheng}}]{Liang10}%
  \BibitemOpen
  \bibfield  {author} {\bibinfo {author} {\bibfnamefont {B.}~\bibnamefont
  {Liang}}, \bibinfo {author} {\bibfnamefont {X.~S.}\ \bibnamefont {Guo}},
  \bibinfo {author} {\bibfnamefont {J.}~\bibnamefont {Tu}}, \bibinfo {author}
  {\bibfnamefont {D.}~\bibnamefont {Zhang}},\ and\ \bibinfo {author}
  {\bibfnamefont {J.~C.}\ \bibnamefont {Cheng}},\ }\href@noop {} {\bibfield
  {journal} {\bibinfo  {journal} {Nature. Mater.}\ }\textbf {\bibinfo {volume}
  {9}},\ \bibinfo {pages} {989} (\bibinfo {year} {2010})}\BibitemShut {NoStop}%
\bibitem [{\citenamefont {Zhu}\ \emph {et~al.}(2004)\citenamefont {Zhu},
  \citenamefont {Dreyer}, \citenamefont {Liebler}, \citenamefont {Riedlinger},
  \citenamefont {Preminger},\ and\ \citenamefont {Zhong}}]{Zhu}%
  \BibitemOpen
  \bibfield  {author} {\bibinfo {author} {\bibfnamefont {S.}~\bibnamefont
  {Zhu}}, \bibinfo {author} {\bibfnamefont {T.}~\bibnamefont {Dreyer}},
  \bibinfo {author} {\bibfnamefont {M.}~\bibnamefont {Liebler}}, \bibinfo
  {author} {\bibfnamefont {R.}~\bibnamefont {Riedlinger}}, \bibinfo {author}
  {\bibfnamefont {G.~M.}\ \bibnamefont {Preminger}},\ and\ \bibinfo {author}
  {\bibfnamefont {P.}~\bibnamefont {Zhong}},\ }\href@noop {} {\bibfield
  {journal} {\bibinfo  {journal} {Ultrasound Med. Biol.}\ }\textbf {\bibinfo
  {volume} {30}},\ \bibinfo {pages} {675 } (\bibinfo {year}
  {2004})}\BibitemShut {NoStop}%
\bibitem [{\citenamefont {Barzanjeh}\ \emph {et~al.}(2017)\citenamefont
  {Barzanjeh}, \citenamefont {Wulf}, \citenamefont {Peruzzo}, \citenamefont
  {Kalaee}, \citenamefont {Dieterle}, \citenamefont {Painter},\ and\
  \citenamefont {Fink}}]{Barzanjeh}%
  \BibitemOpen
  \bibfield  {author} {\bibinfo {author} {\bibfnamefont {S.}~\bibnamefont
  {Barzanjeh}}, \bibinfo {author} {\bibfnamefont {M.}~\bibnamefont {Wulf}},
  \bibinfo {author} {\bibfnamefont {M.}~\bibnamefont {Peruzzo}}, \bibinfo
  {author} {\bibfnamefont {M.}~\bibnamefont {Kalaee}}, \bibinfo {author}
  {\bibfnamefont {P.~B.}\ \bibnamefont {Dieterle}}, \bibinfo {author}
  {\bibfnamefont {O.}~\bibnamefont {Painter}},\ and\ \bibinfo {author}
  {\bibfnamefont {J.~M.}\ \bibnamefont {Fink}},\ }\href@noop {} {\bibfield
  {journal} {\bibinfo  {journal} {Nat. Commun.}\ }\textbf {\bibinfo {volume}
  {8}},\ \bibinfo {pages} {953} (\bibinfo {year} {2017})}\BibitemShut {NoStop}%
\bibitem [{\citenamefont {Yang}\ \emph {et~al.}(2015)\citenamefont {Yang},
  \citenamefont {Gao}, \citenamefont {Shi}, \citenamefont {Lin}, \citenamefont
  {Gao}, \citenamefont {Chong},\ and\ \citenamefont {Zhang}}]{Yang}%
  \BibitemOpen
  \bibfield  {author} {\bibinfo {author} {\bibfnamefont {Z.}~\bibnamefont
  {Yang}}, \bibinfo {author} {\bibfnamefont {F.}~\bibnamefont {Gao}}, \bibinfo
  {author} {\bibfnamefont {X.}~\bibnamefont {Shi}}, \bibinfo {author}
  {\bibfnamefont {X.}~\bibnamefont {Lin}}, \bibinfo {author} {\bibfnamefont
  {Z.}~\bibnamefont {Gao}}, \bibinfo {author} {\bibfnamefont {Y.}~\bibnamefont
  {Chong}},\ and\ \bibinfo {author} {\bibfnamefont {B.}~\bibnamefont {Zhang}},\
  }\bibfield  {title} {\bibinfo {title} {Topological acoustics},\ }\href@noop
  {} {\bibfield  {journal} {\bibinfo  {journal} {Phys. Rev. Lett.}\ }\textbf
  {\bibinfo {volume} {114}},\ \bibinfo {pages} {114301} (\bibinfo {year}
  {2015})}\BibitemShut {NoStop}%
\bibitem [{\citenamefont {Ni}\ \emph {et~al.}(2015)\citenamefont {Ni},
  \citenamefont {He}, \citenamefont {Sun}, \citenamefont {Liu}, \citenamefont
  {Lu}, \citenamefont {Feng},\ and\ \citenamefont {Chen}}]{Ni}%
  \BibitemOpen
  \bibfield  {author} {\bibinfo {author} {\bibfnamefont {X.}~\bibnamefont
  {Ni}}, \bibinfo {author} {\bibfnamefont {C.}~\bibnamefont {He}}, \bibinfo
  {author} {\bibfnamefont {X.-C.}\ \bibnamefont {Sun}}, \bibinfo {author}
  {\bibfnamefont {X.-P.}\ \bibnamefont {Liu}}, \bibinfo {author} {\bibfnamefont
  {M.-H.}\ \bibnamefont {Lu}}, \bibinfo {author} {\bibfnamefont
  {L.}~\bibnamefont {Feng}},\ and\ \bibinfo {author} {\bibfnamefont {Y.-F.}\
  \bibnamefont {Chen}},\ }\href@noop {} {\bibfield  {journal} {\bibinfo
  {journal} {New J. Phys.}\ }\textbf {\bibinfo {volume} {17}},\ \bibinfo
  {pages} {053016} (\bibinfo {year} {2015})}\BibitemShut {NoStop}%
\bibitem [{\citenamefont {Liu}\ and\ \citenamefont
  {Semperlotti}(2018)}]{TWLiu}%
  \BibitemOpen
  \bibfield  {author} {\bibinfo {author} {\bibfnamefont {T.-W.}\ \bibnamefont
  {Liu}}\ and\ \bibinfo {author} {\bibfnamefont {F.}~\bibnamefont
  {Semperlotti}},\ }\href@noop {} {\bibfield  {journal} {\bibinfo  {journal}
  {Phys. Rev. Applied}\ }\textbf {\bibinfo {volume} {9}},\ \bibinfo {pages}
  {014001} (\bibinfo {year} {2018})}\BibitemShut {NoStop}%
\bibitem [{\citenamefont {Li}\ \emph {et~al.}(2013)\citenamefont {Li},
  \citenamefont {Liang}, \citenamefont {Gu}, \citenamefont {Zou},\ and\
  \citenamefont {Cheng}}]{LiY}%
  \BibitemOpen
  \bibfield  {author} {\bibinfo {author} {\bibfnamefont {Y.}~\bibnamefont
  {Li}}, \bibinfo {author} {\bibfnamefont {B.}~\bibnamefont {Liang}}, \bibinfo
  {author} {\bibfnamefont {Z.-m.}\ \bibnamefont {Gu}}, \bibinfo {author}
  {\bibfnamefont {X.-y.}\ \bibnamefont {Zou}},\ and\ \bibinfo {author}
  {\bibfnamefont {J.-c.}\ \bibnamefont {Cheng}},\ }\href
  {https://doi.org/10.1063/1.4817249} {\bibfield  {journal} {\bibinfo
  {journal} {Appl. Phys. Lett.}\ }\textbf {\bibinfo {volume} {103}},\ \bibinfo
  {pages} {053505} (\bibinfo {year} {2013})}\BibitemShut {NoStop}%
\bibitem [{\citenamefont {Yazaki}\ \emph {et~al.}(1998)\citenamefont {Yazaki},
  \citenamefont {Iwata}, \citenamefont {Maekawa},\ and\ \citenamefont
  {Tominaga}}]{Yazaki}%
  \BibitemOpen
  \bibfield  {author} {\bibinfo {author} {\bibfnamefont {T.}~\bibnamefont
  {Yazaki}}, \bibinfo {author} {\bibfnamefont {A.}~\bibnamefont {Iwata}},
  \bibinfo {author} {\bibfnamefont {T.}~\bibnamefont {Maekawa}},\ and\ \bibinfo
  {author} {\bibfnamefont {A.}~\bibnamefont {Tominaga}},\ }\href@noop {}
  {\bibfield  {journal} {\bibinfo  {journal} {Phys. Rev. Lett.}\ }\textbf
  {\bibinfo {volume} {81}},\ \bibinfo {pages} {3128 } (\bibinfo {year}
  {1998})}\BibitemShut {NoStop}%
\bibitem [{\citenamefont {Swift}(1988)}]{Swift88}%
  \BibitemOpen
  \bibfield  {author} {\bibinfo {author} {\bibfnamefont {G.}~\bibnamefont
  {Swift}},\ }\href@noop {} {\bibfield  {journal} {\bibinfo  {journal} {J.
  Acoust. Soc. Am.}\ }\textbf {\bibinfo {volume} {84}},\ \bibinfo {pages} {1145
  } (\bibinfo {year} {1988})}\BibitemShut {NoStop}%
\bibitem [{\citenamefont {Gupta}\ \emph {et~al.}(2017)\citenamefont {Gupta},
  \citenamefont {Lodato},\ and\ \citenamefont {Scalo}}]{Gupta}%
  \BibitemOpen
  \bibfield  {author} {\bibinfo {author} {\bibfnamefont {P.}~\bibnamefont
  {Gupta}}, \bibinfo {author} {\bibfnamefont {G.}~\bibnamefont {Lodato}},\ and\
  \bibinfo {author} {\bibfnamefont {C.}~\bibnamefont {Scalo}},\ }\href@noop {}
  {\bibfield  {journal} {\bibinfo  {journal} {J. Fluid. Mech.}\ }\textbf
  {\bibinfo {volume} {831}},\ \bibinfo {pages} {358 } (\bibinfo {year}
  {2017})}\BibitemShut {NoStop}%
\bibitem [{\citenamefont {Hao}\ \emph {et~al.}(2019)\citenamefont {Hao},
  \citenamefont {Scalo},\ and\ \citenamefont {Semperlotti}}]{HaoJSV19}%
  \BibitemOpen
  \bibfield  {author} {\bibinfo {author} {\bibfnamefont {H.}~\bibnamefont
  {Hao}}, \bibinfo {author} {\bibfnamefont {C.}~\bibnamefont {Scalo}},\ and\
  \bibinfo {author} {\bibfnamefont {F.}~\bibnamefont {Semperlotti}},\
  }\href@noop {} {\bibfield  {journal} {\bibinfo  {journal} {J. Sound. Vib.}\
  }\textbf {\bibinfo {volume} {449}},\ \bibinfo {pages} {30 } (\bibinfo {year}
  {2019})}\BibitemShut {NoStop}%
\bibitem [{\citenamefont {Biwa}\ \emph {et~al.}(2016)\citenamefont {Biwa},
  \citenamefont {Nakamura},\ and\ \citenamefont {Hyodo}}]{Biwa16}%
  \BibitemOpen
  \bibfield  {author} {\bibinfo {author} {\bibfnamefont {T.}~\bibnamefont
  {Biwa}}, \bibinfo {author} {\bibfnamefont {H.}~\bibnamefont {Nakamura}},\
  and\ \bibinfo {author} {\bibfnamefont {H.}~\bibnamefont {Hyodo}},\
  }\href@noop {} {\bibfield  {journal} {\bibinfo  {journal} {Phys. Rev.
  Applied}\ }\textbf {\bibinfo {volume} {5}},\ \bibinfo {pages} {064012}
  (\bibinfo {year} {2016})}\BibitemShut {NoStop}%
\bibitem [{\citenamefont {Senga}\ and\ \citenamefont {Hasegawa}(2016)}]{Senga}%
  \BibitemOpen
  \bibfield  {author} {\bibinfo {author} {\bibfnamefont {M.}~\bibnamefont
  {Senga}}\ and\ \bibinfo {author} {\bibfnamefont {S.}~\bibnamefont
  {Hasegawa}},\ }\href@noop {} {\bibfield  {journal} {\bibinfo  {journal} {J.
  Appl. Phys.}\ }\textbf {\bibinfo {volume} {119}},\ \bibinfo {pages} {204906}
  (\bibinfo {year} {2016})}\BibitemShut {NoStop}%
\bibitem [{\citenamefont {Swift}(2017)}]{SwiftBook}%
  \BibitemOpen
  \bibfield  {author} {\bibinfo {author} {\bibfnamefont {G.~W.}\ \bibnamefont
  {Swift}},\ }\href@noop {} {\emph {\bibinfo {title} {Thermoacoustics A
  Unifying Perspective for Some Engines and Refrigerators}}},\ \bibinfo
  {edition} {2nd}\ ed.\ (\bibinfo {year} {2017})\BibitemShut {NoStop}%
\bibitem [{\citenamefont {Scalo}\ \emph {et~al.}(2015)\citenamefont {Scalo},
  \citenamefont {Lele},\ and\ \citenamefont {Hesselink}}]{Scalo}%
  \BibitemOpen
  \bibfield  {author} {\bibinfo {author} {\bibfnamefont {C.}~\bibnamefont
  {Scalo}}, \bibinfo {author} {\bibfnamefont {S.}~\bibnamefont {Lele}},\ and\
  \bibinfo {author} {\bibfnamefont {L.}~\bibnamefont {Hesselink}},\ }\href@noop
  {} {\bibfield  {journal} {\bibinfo  {journal} {J. Fluid. Mech.}\ }\textbf
  {\bibinfo {volume} {766}},\ \bibinfo {pages} {368 } (\bibinfo {year}
  {2015})}\BibitemShut {NoStop}%
\bibitem [{\citenamefont {Chen}\ \emph {et~al.}(2018)\citenamefont {Chen},
  \citenamefont {Tang},\ and\ \citenamefont {Mace}}]{Chen0}%
  \BibitemOpen
  \bibfield  {author} {\bibinfo {author} {\bibfnamefont {G.}~\bibnamefont
  {Chen}}, \bibinfo {author} {\bibfnamefont {L.}~\bibnamefont {Tang}},\ and\
  \bibinfo {author} {\bibfnamefont {B.}~\bibnamefont {Mace}},\ }\href@noop {}
  {\bibfield  {journal} {\bibinfo  {journal} {Int. J. Heat Mass Transf.}\
  }\textbf {\bibinfo {volume} {123}},\ \bibinfo {pages} {367 } (\bibinfo {year}
  {2018})}\BibitemShut {NoStop}%
\bibitem [{\citenamefont {Hao}\ \emph {et~al.}(2018)\citenamefont {Hao},
  \citenamefont {Scalo}, \citenamefont {Sen},\ and\ \citenamefont
  {Semperlotti}}]{HaoJAP}%
  \BibitemOpen
  \bibfield  {author} {\bibinfo {author} {\bibfnamefont {H.}~\bibnamefont
  {Hao}}, \bibinfo {author} {\bibfnamefont {C.}~\bibnamefont {Scalo}}, \bibinfo
  {author} {\bibfnamefont {M.}~\bibnamefont {Sen}},\ and\ \bibinfo {author}
  {\bibfnamefont {F.}~\bibnamefont {Semperlotti}},\ }\href@noop {} {\bibfield
  {journal} {\bibinfo  {journal} {J. Appl. Phys.}\ }\textbf {\bibinfo {volume}
  {123}},\ \bibinfo {pages} {024903} (\bibinfo {year} {2018})}\BibitemShut
  {NoStop}%
\bibitem [{\citenamefont {Wang}\ \emph {et~al.}(2015)\citenamefont {Wang},
  \citenamefont {Wang},\ and\ \citenamefont {Laude}}]{Wang15}%
  \BibitemOpen
  \bibfield  {author} {\bibinfo {author} {\bibfnamefont {Y.-F.}\ \bibnamefont
  {Wang}}, \bibinfo {author} {\bibfnamefont {Y.-S.}\ \bibnamefont {Wang}},\
  and\ \bibinfo {author} {\bibfnamefont {V.}~\bibnamefont {Laude}},\
  }\href@noop {} {\bibfield  {journal} {\bibinfo  {journal} {Phys. Rev. B}\
  }\textbf {\bibinfo {volume} {92}},\ \bibinfo {pages} {104110} (\bibinfo
  {year} {2015})}\BibitemShut {NoStop}%
\bibitem [{\citenamefont {Liu}\ \emph {et~al.}(2008)\citenamefont {Liu},
  \citenamefont {Yu}, \citenamefont {Zhao}, \citenamefont {Wen},\ and\
  \citenamefont {Wen}}]{Liu}%
  \BibitemOpen
  \bibfield  {author} {\bibinfo {author} {\bibfnamefont {Y.}~\bibnamefont
  {Liu}}, \bibinfo {author} {\bibfnamefont {D.}~\bibnamefont {Yu}}, \bibinfo
  {author} {\bibfnamefont {H.}~\bibnamefont {Zhao}}, \bibinfo {author}
  {\bibfnamefont {J.}~\bibnamefont {Wen}},\ and\ \bibinfo {author}
  {\bibfnamefont {X.}~\bibnamefont {Wen}},\ }\href
  {https://doi.org/10.1088/0022-3727/41/6/065503} {\bibfield  {journal}
  {\bibinfo  {journal} {J. Phys. D Appl. Phys.}\ }\textbf {\bibinfo {volume}
  {41}},\ \bibinfo {pages} {065503} (\bibinfo {year} {2008})}\BibitemShut
  {NoStop}%
\bibitem [{\citenamefont {Hwan~Oh}\ \emph {et~al.}(2013)\citenamefont
  {Hwan~Oh}, \citenamefont {Jae~Kim},\ and\ \citenamefont {Young~Kim}}]{Oh}%
  \BibitemOpen
  \bibfield  {author} {\bibinfo {author} {\bibfnamefont {J.}~\bibnamefont
  {Hwan~Oh}}, \bibinfo {author} {\bibfnamefont {Y.}~\bibnamefont {Jae~Kim}},\
  and\ \bibinfo {author} {\bibfnamefont {Y.}~\bibnamefont {Young~Kim}},\
  }\href@noop {} {\bibfield  {journal} {\bibinfo  {journal} {J. Appl. Phys.}\
  }\textbf {\bibinfo {volume} {113}},\ \bibinfo {pages} {106101} (\bibinfo
  {year} {2013})}\BibitemShut {NoStop}%
\bibitem [{\citenamefont {Theocharis}\ \emph {et~al.}(2014)\citenamefont
  {Theocharis}, \citenamefont {Richoux}, \citenamefont {García}, \citenamefont
  {Merkel},\ and\ \citenamefont {Tournat}}]{Theocharis}%
  \BibitemOpen
  \bibfield  {author} {\bibinfo {author} {\bibfnamefont {G.}~\bibnamefont
  {Theocharis}}, \bibinfo {author} {\bibfnamefont {O.}~\bibnamefont {Richoux}},
  \bibinfo {author} {\bibfnamefont {V.~R.}\ \bibnamefont {García}}, \bibinfo
  {author} {\bibfnamefont {A.}~\bibnamefont {Merkel}},\ and\ \bibinfo {author}
  {\bibfnamefont {V.}~\bibnamefont {Tournat}},\ }\href@noop {} {\bibfield
  {journal} {\bibinfo  {journal} {New J. Phys.}\ }\textbf {\bibinfo {volume}
  {16}} (\bibinfo {year} {2014})}\BibitemShut {NoStop}%
\bibitem [{\citenamefont {Henr\'{\i}quez}\ \emph {et~al.}(2017)\citenamefont
  {Henr\'{\i}quez}, \citenamefont {Garc\'{\i}a-Chocano},\ and\ \citenamefont
  {S\'anchez-Dehesa}}]{Henriquez}%
  \BibitemOpen
  \bibfield  {author} {\bibinfo {author} {\bibfnamefont {V.~C.}\ \bibnamefont
  {Henr\'{\i}quez}}, \bibinfo {author} {\bibfnamefont {V.~M.}\ \bibnamefont
  {Garc\'{\i}a-Chocano}},\ and\ \bibinfo {author} {\bibfnamefont
  {J.}~\bibnamefont {S\'anchez-Dehesa}},\ }\href
  {10.1103/PhysRevApplied.8.014029} {\bibfield  {journal} {\bibinfo  {journal}
  {Phys. Rev. Applied}\ }\textbf {\bibinfo {volume} {8}},\ \bibinfo {pages}
  {014029} (\bibinfo {year} {2017})}\BibitemShut {NoStop}%
\bibitem [{\citenamefont {Rott}(1969)}]{Rott}%
  \BibitemOpen
  \bibfield  {author} {\bibinfo {author} {\bibfnamefont {N.}~\bibnamefont
  {Rott}},\ }\href@noop {} {\bibfield  {journal} {\bibinfo  {journal} {Z.
  Angew. Math. Phys.}\ }\textbf {\bibinfo {volume} {20}},\ \bibinfo {pages}
  {230 } (\bibinfo {year} {1969})}\BibitemShut {NoStop}%
\bibitem [{\citenamefont {Guedra}\ and\ \citenamefont
  {Penelet}(2012)}]{Penelet}%
  \BibitemOpen
  \bibfield  {author} {\bibinfo {author} {\bibfnamefont {M.}~\bibnamefont
  {Guedra}}\ and\ \bibinfo {author} {\bibfnamefont {G.}~\bibnamefont
  {Penelet}},\ }\href@noop {} {\bibfield  {journal} {\bibinfo  {journal} {Acta.
  Acust. united Ac.}\ }\textbf {\bibinfo {volume} {98}},\ \bibinfo {pages} {232
  } (\bibinfo {year} {2012})}\BibitemShut {NoStop}%
\bibitem [{\citenamefont {Lin}\ \emph {et~al.}(2016)\citenamefont {Lin},
  \citenamefont {Scalo},\ and\ \citenamefont {Hesselink}}]{Lin}%
  \BibitemOpen
  \bibfield  {author} {\bibinfo {author} {\bibfnamefont {J.}~\bibnamefont
  {Lin}}, \bibinfo {author} {\bibfnamefont {C.}~\bibnamefont {Scalo}},\ and\
  \bibinfo {author} {\bibfnamefont {L.}~\bibnamefont {Hesselink}},\ }\href@noop
  {} {\bibfield  {journal} {\bibinfo  {journal} {J. Fluid. Mech.}\ }\textbf
  {\bibinfo {volume} {808}},\ \bibinfo {pages} {19–60} (\bibinfo {year}
  {2016})}\BibitemShut {NoStop}%
\bibitem [{\citenamefont {Chen}\ \emph {et~al.}(2019)\citenamefont {Chen},
  \citenamefont {Tang},\ and\ \citenamefont {Mace}}]{ChenATE19}%
  \BibitemOpen
  \bibfield  {author} {\bibinfo {author} {\bibfnamefont {G.}~\bibnamefont
  {Chen}}, \bibinfo {author} {\bibfnamefont {L.}~\bibnamefont {Tang}},\ and\
  \bibinfo {author} {\bibfnamefont {B.~R.}\ \bibnamefont {Mace}},\ }\href@noop
  {} {\bibfield  {journal} {\bibinfo  {journal} {Appl. Therm. Eng.}\ }\textbf
  {\bibinfo {volume} {150}},\ \bibinfo {pages} {532 } (\bibinfo {year}
  {2019})}\BibitemShut {NoStop}%
\bibitem [{\citenamefont {Hao}\ \emph {et~al.}(2020)\citenamefont {Hao},
  \citenamefont {Scalo},\ and\ \citenamefont {Semperlotti}}]{HaoJSV20}%
  \BibitemOpen
  \bibfield  {author} {\bibinfo {author} {\bibfnamefont {H.}~\bibnamefont
  {Hao}}, \bibinfo {author} {\bibfnamefont {C.}~\bibnamefont {Scalo}},\ and\
  \bibinfo {author} {\bibfnamefont {F.}~\bibnamefont {Semperlotti}},\
  }\href@noop {} {\bibfield  {journal} {\bibinfo  {journal} {J. Sound. Vib.}\
  }\textbf {\bibinfo {volume} {470}},\ \bibinfo {pages} {115159} (\bibinfo
  {year} {2020})}\BibitemShut {NoStop}%
\bibitem [{\citenamefont {Hao}\ \emph {et~al.}(2021)\citenamefont {Hao},
  \citenamefont {Scalo},\ and\ \citenamefont {Semperlotti}}]{HaoMSSP}%
  \BibitemOpen
  \bibfield  {author} {\bibinfo {author} {\bibfnamefont {H.}~\bibnamefont
  {Hao}}, \bibinfo {author} {\bibfnamefont {C.}~\bibnamefont {Scalo}},\ and\
  \bibinfo {author} {\bibfnamefont {F.}~\bibnamefont {Semperlotti}},\
  }\href@noop {} {\bibfield  {journal} {\bibinfo  {journal} {Mech. Syst.
  Signal. Process.}\ }\textbf {\bibinfo {volume} {148}},\ \bibinfo {pages}
  {107143} (\bibinfo {year} {2021})}\BibitemShut {NoStop}%
\bibitem [{\citenamefont {Suzuki}\ and\ \citenamefont {Yu}(1998)}]{Suzuki}%
  \BibitemOpen
  \bibfield  {author} {\bibinfo {author} {\bibfnamefont {T.}~\bibnamefont
  {Suzuki}}\ and\ \bibinfo {author} {\bibfnamefont {P.~K.}\ \bibnamefont
  {Yu}},\ }\href@noop {} {\bibfield  {journal} {\bibinfo  {journal} {J. Mech.
  Phys. Solids}\ }\textbf {\bibinfo {volume} {46}},\ \bibinfo {pages} {115 }
  (\bibinfo {year} {1998})}\BibitemShut {NoStop}%
\bibitem [{\citenamefont {Tan}\ \emph {et~al.}(2020)\citenamefont {Tan},
  \citenamefont {Wei},\ and\ \citenamefont {Jin}}]{Tan}%
  \BibitemOpen
  \bibfield  {author} {\bibinfo {author} {\bibfnamefont {J.}~\bibnamefont
  {Tan}}, \bibinfo {author} {\bibfnamefont {J.}~\bibnamefont {Wei}},\ and\
  \bibinfo {author} {\bibfnamefont {T.}~\bibnamefont {Jin}},\ }\href@noop {}
  {\bibfield  {journal} {\bibinfo  {journal} {Appl. Energy}\ }\textbf {\bibinfo
  {volume} {262}},\ \bibinfo {pages} {114539} (\bibinfo {year}
  {2020})}\BibitemShut {NoStop}%
\bibitem [{\citenamefont {Shi}\ \emph {et~al.}(2017)\citenamefont {Shi},
  \citenamefont {Dubois}, \citenamefont {Wang},\ and\ \citenamefont
  {Zhang}}]{Shi}%
  \BibitemOpen
  \bibfield  {author} {\bibinfo {author} {\bibfnamefont {C.}~\bibnamefont
  {Shi}}, \bibinfo {author} {\bibfnamefont {M.}~\bibnamefont {Dubois}},
  \bibinfo {author} {\bibfnamefont {Y.}~\bibnamefont {Wang}},\ and\ \bibinfo
  {author} {\bibfnamefont {X.}~\bibnamefont {Zhang}},\ }\href
  {https://doi.org/10.1073/pnas.1704450114} {\bibfield  {journal} {\bibinfo
  {journal} {Proc. Natl. Acad. Sci.}\ }\textbf {\bibinfo {volume} {114}},\
  \bibinfo {pages} {7250} (\bibinfo {year} {2017})}\BibitemShut {NoStop}%
\bibitem [{\citenamefont {Bradley}(1994)}]{BradleyTR}%
  \BibitemOpen
  \bibfield  {author} {\bibinfo {author} {\bibfnamefont {C.~E.}\ \bibnamefont
  {Bradley}},\ }\href@noop {} {\emph {\bibinfo {title} {Linear and nonlinear
  acoustic Bloch wave propagation in periodic waveguides}}},\ \bibinfo {type}
  {Tech. Rep.}\ \bibinfo {number} {ARL-TR-94-10}\ (\bibinfo  {institution} {The
  University of Texas at Austin},\ \bibinfo {year} {1994})\BibitemShut
  {NoStop}%
\bibitem [{\citenamefont {Jin}\ \emph {et~al.}(2016)\citenamefont {Jin},
  \citenamefont {Yang}, \citenamefont {Wang}, \citenamefont {Liu},\ and\
  \citenamefont {Feng}}]{Jin}%
  \BibitemOpen
  \bibfield  {author} {\bibinfo {author} {\bibfnamefont {T.}~\bibnamefont
  {Jin}}, \bibinfo {author} {\bibfnamefont {R.}~\bibnamefont {Yang}}, \bibinfo
  {author} {\bibfnamefont {Y.}~\bibnamefont {Wang}}, \bibinfo {author}
  {\bibfnamefont {Y.}~\bibnamefont {Liu}},\ and\ \bibinfo {author}
  {\bibfnamefont {Y.}~\bibnamefont {Feng}},\ }\href@noop {} {\bibfield
  {journal} {\bibinfo  {journal} {Appl. Energy}\ }\textbf {\bibinfo {volume}
  {183}},\ \bibinfo {pages} {290 } (\bibinfo {year} {2016})}\BibitemShut
  {NoStop}%
\bibitem [{\citenamefont {de~Blok}(2008)}]{deBlok}%
  \BibitemOpen
  \bibfield  {author} {\bibinfo {author} {\bibfnamefont {K.}~\bibnamefont
  {de~Blok}},\ }\href@noop {} {\bibfield  {journal} {\bibinfo  {journal} {J.
  Acoust. Soc. Am.}\ }\textbf {\bibinfo {volume} {123}},\ \bibinfo {pages}
  {3541} (\bibinfo {year} {2008})}\BibitemShut {NoStop}%
\bibitem [{\citenamefont {Dubois}\ \emph {et~al.}(2017)\citenamefont {Dubois},
  \citenamefont {Shi}, \citenamefont {Zhu}, \citenamefont {Wang},\ and\
  \citenamefont {Zhang}}]{Dubois}%
  \BibitemOpen
  \bibfield  {author} {\bibinfo {author} {\bibfnamefont {M.}~\bibnamefont
  {Dubois}}, \bibinfo {author} {\bibfnamefont {C.}~\bibnamefont {Shi}},
  \bibinfo {author} {\bibfnamefont {X.}~\bibnamefont {Zhu}}, \bibinfo {author}
  {\bibfnamefont {Y.}~\bibnamefont {Wang}},\ and\ \bibinfo {author}
  {\bibfnamefont {X.}~\bibnamefont {Zhang}},\ }\href@noop {} {\bibfield
  {journal} {\bibinfo  {journal} {Nat. Commun.}\ }\textbf {\bibinfo {volume}
  {8}},\ \bibinfo {pages} {14871} (\bibinfo {year} {2017})}\BibitemShut
  {NoStop}%
\bibitem [{\citenamefont {Zhu}(2013)}]{ZhuXF}%
  \BibitemOpen
  \bibfield  {author} {\bibinfo {author} {\bibfnamefont {X.-F.}\ \bibnamefont
  {Zhu}},\ }\href
  {https://doi.org/https://doi.org/10.1016/j.physleta.2013.05.038} {\bibfield
  {journal} {\bibinfo  {journal} {Phys. Lett. A}\ }\textbf {\bibinfo {volume}
  {377}},\ \bibinfo {pages} {1784 } (\bibinfo {year} {2013})}\BibitemShut
  {NoStop}%
\bibitem [{\citenamefont {Zhu}\ and\ \citenamefont {Semperlotti}(2017)}]{ZhuH}%
  \BibitemOpen
  \bibfield  {author} {\bibinfo {author} {\bibfnamefont {H.}~\bibnamefont
  {Zhu}}\ and\ \bibinfo {author} {\bibfnamefont {F.}~\bibnamefont
  {Semperlotti}},\ }\href {https://doi.org/10.1103/PhysRevApplied.8.064031}
  {\bibfield  {journal} {\bibinfo  {journal} {Phys. Rev. Applied}\ }\textbf
  {\bibinfo {volume} {8}},\ \bibinfo {pages} {064031} (\bibinfo {year}
  {2017})}\BibitemShut {NoStop}%
\bibitem [{\citenamefont {Kaina}\ \emph {et~al.}(2015)\citenamefont {Kaina},
  \citenamefont {Lemoult}, \citenamefont {Fink},\ and\ \citenamefont
  {Lerosey}}]{Kaina}%
  \BibitemOpen
  \bibfield  {author} {\bibinfo {author} {\bibfnamefont {N.}~\bibnamefont
  {Kaina}}, \bibinfo {author} {\bibfnamefont {F.}~\bibnamefont {Lemoult}},
  \bibinfo {author} {\bibfnamefont {M.}~\bibnamefont {Fink}},\ and\ \bibinfo
  {author} {\bibfnamefont {G.}~\bibnamefont {Lerosey}},\ }\href@noop {}
  {\bibfield  {journal} {\bibinfo  {journal} {Nature}\ }\textbf {\bibinfo
  {volume} {525}},\ \bibinfo {pages} {77} (\bibinfo {year} {2015})}\BibitemShut
  {NoStop}%
\bibitem [{\citenamefont {Lee}\ and\ \citenamefont {Wright}(2016)}]{Lee}%
  \BibitemOpen
  \bibfield  {author} {\bibinfo {author} {\bibfnamefont {S.~H.}\ \bibnamefont
  {Lee}}\ and\ \bibinfo {author} {\bibfnamefont {O.~B.}\ \bibnamefont
  {Wright}},\ }\href {https://doi.org/10.1103/PhysRevB.93.024302} {\bibfield
  {journal} {\bibinfo  {journal} {Phys. Rev. B}\ }\textbf {\bibinfo {volume}
  {93}},\ \bibinfo {pages} {024302} (\bibinfo {year} {2016})}\BibitemShut
  {NoStop}%
\bibitem [{\citenamefont {Ceperley}(1979)}]{Ceperley}%
  \BibitemOpen
  \bibfield  {author} {\bibinfo {author} {\bibfnamefont {P.}~\bibnamefont
  {Ceperley}},\ }\href@noop {} {\bibfield  {journal} {\bibinfo  {journal} {J.
  Acoust. Soc. Am.}\ }\textbf {\bibinfo {volume} {66}},\ \bibinfo {pages} {1508
  } (\bibinfo {year} {1979})}\BibitemShut {NoStop}%
\end{thebibliography}%

\end{document}